\documentclass[preprint,aps]{revtex4}
\usepackage[latin9]{luainputenc}
\usepackage{float}
\usepackage{amsmath}
\usepackage{amssymb}
\usepackage{graphicx}
\usepackage{setspace}
\usepackage[unicode=true,pdfusetitle,
bookmarks=true,bookmarksnumbered=false,bookmarksopen=false,
breaklinks=false,pdfborder={0 0 1},backref=false,colorlinks=false]
{hyperref}

\makeatletter

\@ifundefined{textcolor}{}
{%
	\definecolor{BLACK}{gray}{0}
	\definecolor{WHITE}{gray}{1}
	\definecolor{RED}{rgb}{1,0,0}
	\definecolor{GREEN}{rgb}{0,1,0}
	\definecolor{BLUE}{rgb}{0,0,1}
	\definecolor{CYAN}{cmyk}{1,0,0,0}
	\definecolor{MAGENTA}{cmyk}{0,1,0,0}
	\definecolor{YELLOW}{cmyk}{0,0,1,0}
}

\usepackage{amsfonts}
\usepackage{bm}
\usepackage[dvips]{epsfig}
\usepackage[dvips]{graphicx}
\usepackage{float}
\usepackage{dcolumn}
\usepackage{ifpdf}
\usepackage{dcolumn}
\usepackage{multirow}

\newcommand{\be}{\begin{equation}}
\newcommand{\ee}{\end{equation}}
\newcommand{\bes}{\begin{subequations}}
	\newcommand{\ees}{\end{subequations}}
\newcommand{\ben}{\begin{eqnarray}}
\newcommand{\een}{\end{eqnarray}}

\makeatother

\begin{document}
\title{Solitary oscillations and multiple antikink-kink pairs in the double sine-Gordon model}
 \author{Fabiano C. Simas$^{1,2}$,  Fred C. Lima$^{2}$, K. Z. Nobrega$^{3,4}$, Adalto R. Gomes$^{2}$}
 \email{argomes.ufma@gmail.com, adalto.gomes@ufma.br}
  \noaffiliation
\affiliation{
$^1$ Centro de Ci\^encias Agr\'arias e Ambientais-CCAA, Universidade Federal do Maranh\~ao
(UFMA), 65500-000, Chapadinha, Maranh\~ao, Brazil\\
$^2$ Programa de P\'os-Gradua\c c\~ao em F\'\i sica, Universidade Federal do Maranh\~ao (UFMA),\\
Campus Universit\'ario do Bacanga, 65085-580, S\~ao Lu\'\i s, Maranh\~ao, Brazil\\
$^3$ Departamento de Eletro-Eletr\^onica, Instituto Federal de Educa\c c\~ao, Ci\^encia e
Tecnologia do Maranh\~ao (IFMA), Campus Monte Castelo, 65030-005, S\~ao Lu\'is, Maranh\~ao, Brazil\\
$^4$ Departamento de Engenharia Teleinform\'atica, Universidade Federal do Cear\'a (UFC), 60455-640, Fortaleza, Cear\'a, Brazil 
}
\noaffiliation

\begin{abstract}
We study kink-antikink collisions in a particular case of the double sine-Gordon model depending on only one parameter $r$.  The scattering process of large kink-antikink shows the changing of the topological sector. For some parameter intervals we observed two connected effects: the production of up to five antikink-kink pairs and up to three solitary oscillations.  The scattering process for small kink-antikink has several possibilities: the changing of the topological sector, one-bounce collision, two-bounce collision, or formation of a bion state. In particular, we observed for small values of $r$ and velocities, the formation of false two-bounce windows and the suppression of true two-bounce windows, despite the presence of an internal shape mode. 
\end{abstract}



\maketitle


\section{ Introduction }


Domain walls in $(3,1)$ dimensions and kinks/antikinks in $(1,1)$ dimensions are solutions of nonlinear field theories.  Due to their stability and topological structure, these spatially localized configurations can propagate freely without losing their shape \cite{daux, vacha}. Domain walls and kinks have  applications in several areas of science including early universe cosmology \cite{agui,gly, class}, pulsar glitches \cite{yasui}, ferroelectrics \cite{stru},  optical fibers \cite{molle} and DNA \cite{yaku1}.  In integrable models like the sine-Gordon, the structure of kink-antikink scattering is simple, presenting a totally elastic behavior with at most a phase shift.  On the other hand, nonintegrable models show a richer pattern of kink-antikink scattering. The investigation of new patterns of kink scattering is interesting to understand some aspects of nonlinearity connected with physical systems. 

Among nonintegrable models, the $\phi^4$ model is the simplest and by far more studied in the literature (see  \cite{phi4} and references therein). It is known that the kink-antikink ($K \bar K$) collision depends on the initial velocity $v$. There is a critical velocity $v_c$ such that, for $v>v_c$ there is an inelastic scattering between the kink and the antikink. For $v<v_c$ the kink-antikink remain connected and irradiate continuously. For some values of $v$, we can observe the annihilation of the pair and the fractal resonance structure. The structure formed, known as escape windows or two-bounce windows, is related to the resonant energy exchange mechanism between the translational and vibrational modes \cite{csw}. See Ref. \cite{kudry} for a review of early works on this complex structure. 

 More recently, the numerical investigation of other models has been revealed some unexpected aspects of nonlinearity. In particular, in the $\phi^6$ model \cite{dorey2}, despite the absence of vibrational mode for one kink, the structure resonant scattering appears. There, the authors explained the results  considering the perturbation of the whole kink-antikink pair. In the model of Ref. \cite{sgn}, two-bounce windows are suppressed despite the presence of more than one internal mode. This effect was interpreted as a kind of destructive interference between the several modes. In the Refs. \cite{dorey1,sgn2} we can see once again the importance of vibrational mode in the appearance of two-bounce windows.

A large number of models with kink solutions have been subject of investigation, including higher-order polynomial models \cite{demir,weigel1,gani6,aza1,tail,phi8}, models with two scalar fields \cite{hala1,alonso1,alonso2,alonso3}, hyperbolic models \cite{bazeia2,gomes1,gomes2} and multikinks \cite{saad,marja3,gani}. One can also cite studies of interaction of kink with a boundary \cite{dorey3,art,fred},  models with generalized dynamics \cite{adalto3, liu1}, with Lorentz symmetry violation \cite{maeda}, wobbling kinks \cite{wob} and effects of the transition a bound mode through the mass threshold during the scattering process \cite{spec1,spec2,spec3}.

The double sine-Gordon (DSG) model  was explored long ago in the Ref. \cite{kumar1} to describe the spin dynamics in the superfluid ${}^3He$. There, the longitudinal nuclear
magnetic resonance corresponds to the oscillation of an angle $\phi$ in a potential well. In the B phase, the longitudinal solitons in the Leggett configuration (where the symmetry breaking axis is parallel to the external field) results in a modified sine-Gordon equation for $\phi$ with two types of solutions: small soliton and large soliton. Some results on soliton-antisoliton scattering for both small soliton and large soliton pairs were presented in the Refs. \cite{kumar2, kumar3}. For a more gentle introduction on the subject, see the Ref. \cite{dp}.
The DSG admits quantization and applications 
in statistical field theory \cite{delf}, being closely related to the Ashkin-Teller model that describes two planar Ising models interacting through a
local four-spin interaction. Moreover, it is a nice example of a quantum phase transition. This rich phenomenology is addressed considering the kink configurations  of the model and their bound states \cite{delf}.   
A recent paper derived the dynamic equation of molecular motion for  twisted nematic liquid crystal under applied electric an magnetic fields, showing that it takes the form of a DSG model \cite{Li}. The recent observation of a fractional vortex in a superconducting bi-layer \cite{thyna} indicates that kinks can be formed in such systems.  For fractional vortices in superconductors see the Chap. 6 of the Ref. \cite{babaev}. In the Ref. \cite{yht}, kinks in a two-band superconductor are described by the DSG Model.  Pulsar glitches are sudden changes in the rotation frequency of neutron stars \cite{ppls}. This is an evidence of the existence of superconductor states in the core of neutron stars. Collective excitations along a vortex line in neutron ${}^3P_2$ superfluids in neutron stars results, for low energies, in a kink of the DSG model \cite{chm}. The generalized sine-Gordon model has been found in the study of the strong/weak coupling sectors of the $sl(N,\mathbb{C})$ affine Toda model coupled to matter fields \cite{blas1}. The DSG model appears in the reduction to one field of the $sl(3,\mathbb{C})$ generalized sine-Gordon model. In the Ref. \cite{uchi} the DSG model was proposed as an extended hadron model. That work was extended in the Ref \cite{blas1} to multiflavor Dirac fields, such that the DSG kink solution describes a multi-baryon. Then, the DSG spectrum and kink-antikink system can be useful for the description of multiflavor spectrum and some resonances in $QCD_2$ \cite{blas1}. The DSG model also appears in the effective action describing exotic baryons in $QCD_2$ \cite{blas2,blas3}.

The last paragraph illustrated several important applications where the results of kink scattering in the DSG model can be useful. Aspects of kinks in the DSG model such as small oscillations, internal modes, radiation and analytical methods were studied in the Refs. \cite{kDSG1,kDSG2,kDSG3,kDSG4,kDSG5,kDSG6,kDSG7,baz1}. Kink scattering in the DSG model was investigated in the Refs. \cite{dsg1,dsg2,dsg3,dsg4,dsg5,dsg6,dsg7,dsg8,dsg9,dsg10,dsg11,dsg12}. 
Here we consider a particular case of the DSG model, looking for the structure of the kink-antikink scattering.
We found some aspects not reported before, such as the production of up to three solitary oscillations and multiple antikink-kink pairs. We show that these effects are interrelated.

In the Sect. II we discuss the model and two types of kinks: large kinks and small kinks. For large kinks it is shown the production of multiple extra antikink-kink pairs together with the changing of the topological sector. We report the formation of  several oscillations resembling oscillons, but with much shorter lifetimes. For small kinks we investigate the appearance of the structure of two-bounce windows. We present our main conclusion in the Sect. III.


\section { The Model }


We consider the action with standard dynamics in $(1,1)$-dimensions in a Minkowski spacetime

\be
S=\int dt dx { \bigg( \frac12  \partial_{\mu} \partial^{\mu} \phi - V(\phi) \bigg) }. 
\label{action}
\ee
\begin{figure}
	\includegraphics[{angle=0,width=10cm,height=6cm}]{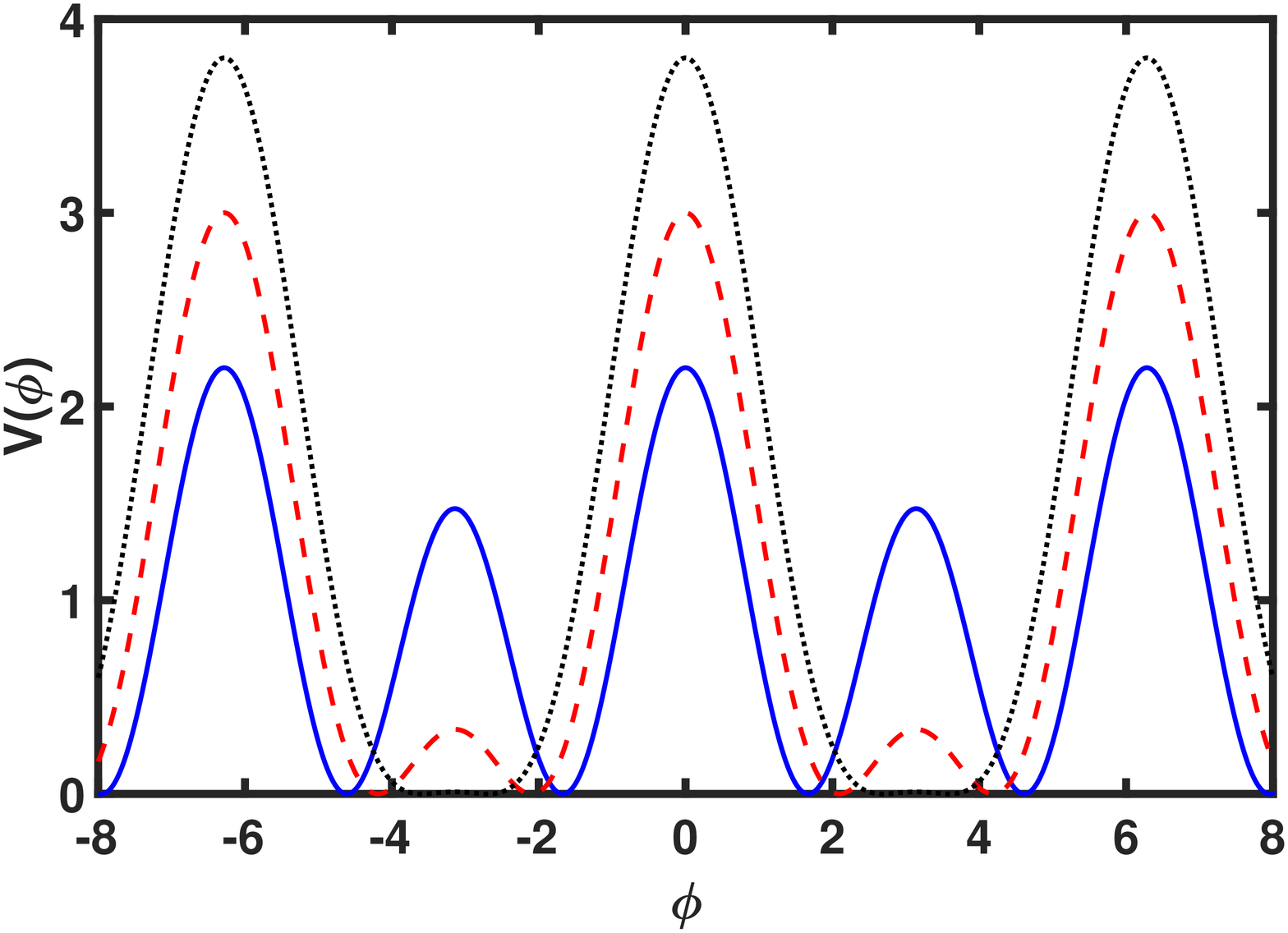}
	\caption{Potential $V(\phi)$ given by the Eq. (\ref{pot}). It  is fixed $r=0.1$ (blue solid), $r=0.5$ (red dash), $r=0.9$ (black dot).}
		\label{fig1}
	\end{figure}

The double sine-Gordon is defined by the potential
\be
V(\phi)=\frac{\lambda^2}{2\alpha^2\nu^2} [ \cos^2(\nu\phi) - \alpha^2\sin^2(\nu\phi) ]^2.
\label{VDSG}
\ee
In this work we consider the potential \cite{baz1}
\be
V(\phi) = \frac{2}{1+r}\big( \cos(\phi)+r \big)^2,
\label{pot}
\ee
where $0<r<1$ is a parameter. This potential is a particular case of the potential (\ref{VDSG}), with \cite{baz1} $\alpha=\sqrt{(1-r)/(1+r)}$, $\nu=1/2$ and $\lambda=\sqrt{1-r}$, and is depicted in Fig. \ref{fig1} for some values of $r$. This potential is periodic, with period $2\pi$. For $-2\pi<\phi<2\pi$ and for the parameter $0<r<1$ the potential contains four minima.
Note from the figure that the minima are separated by large and small barriers. Corresponding topological solutions are named large and small kinks. For $r\to 0$ the potential goes to the integrable sine-Gordon. The growth of $r$ reduces the height for the small kink until the critical value $r\to1$, where the potential goes to zero at $\phi=\pm\pi$. In this limit the potential is very close to zero for a finite interval around $\phi=\pm\pi$, as can be seen in the Fig. \ref{fig1}. This agrees with ${V_\phi}(\phi=\pm\pi)={V_{\phi\phi}}(\phi=\pm\pi)=0$. The main motivation of  the potential given by the Eq. (\ref{pot}) was the dependence of only one parameter: if $r$ is small one can explore the vicinity of the integrable sine-Gordon model \cite{baz1}. Here we will consider this potential in the range $0<r<1$, exploring the main characteristics of scattering of large kinks and small kinks.

The equation of motion is given by

\be
\frac{\partial^2 \phi}{\partial t^2}-\frac{\partial^2 \phi}{\partial x^2}+\frac{d V(\phi)}{d \phi}=0,
\label{eqm}
\ee
and static kinks $\phi_S(x)$ are solutions that connect two sectors of the potential and minimize the energy. Perturbing linearly the scalar field around one static kink solution $\phi_S(x)$ as $\phi(x,t) = \phi_S(x) + \eta(x) \cos(\omega t)$ leads to Schrodinger-like equation 

\be
-\frac{d^2 \eta(x)}{dx^2} + V_{sch}(x) \eta(x) = \omega^2 \eta(x),
\ee
where $V_{sch}(x) = V_{\phi \phi}(\phi_S(x))$. The study of the Schr\"odinger-like potential is useful for understanding some aspects of the scattering structure.

\begin{figure}
	\includegraphics[{angle=0,width=8cm,height=5cm}]{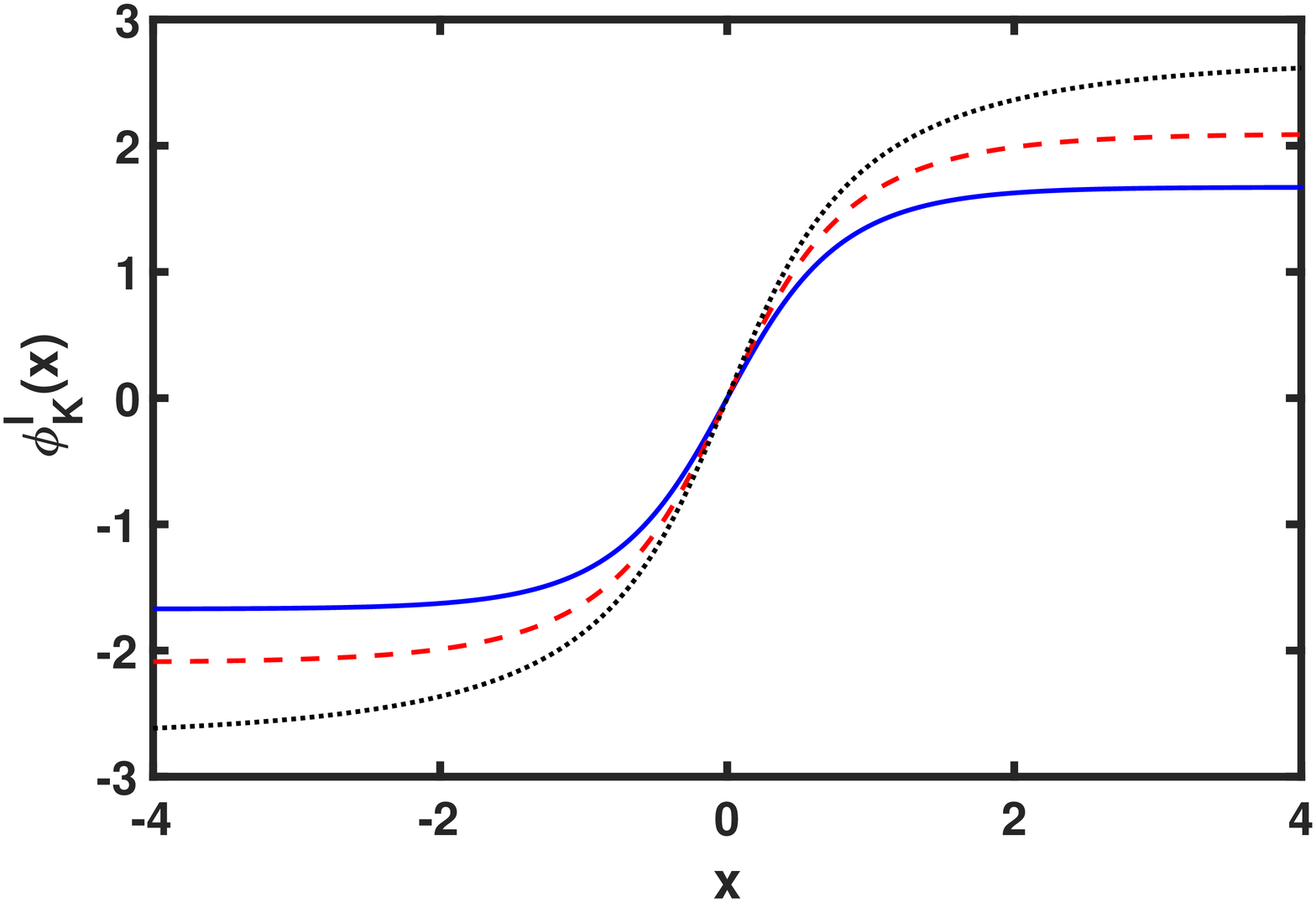}
	\includegraphics[{angle=0,width=8cm,height=5cm}]{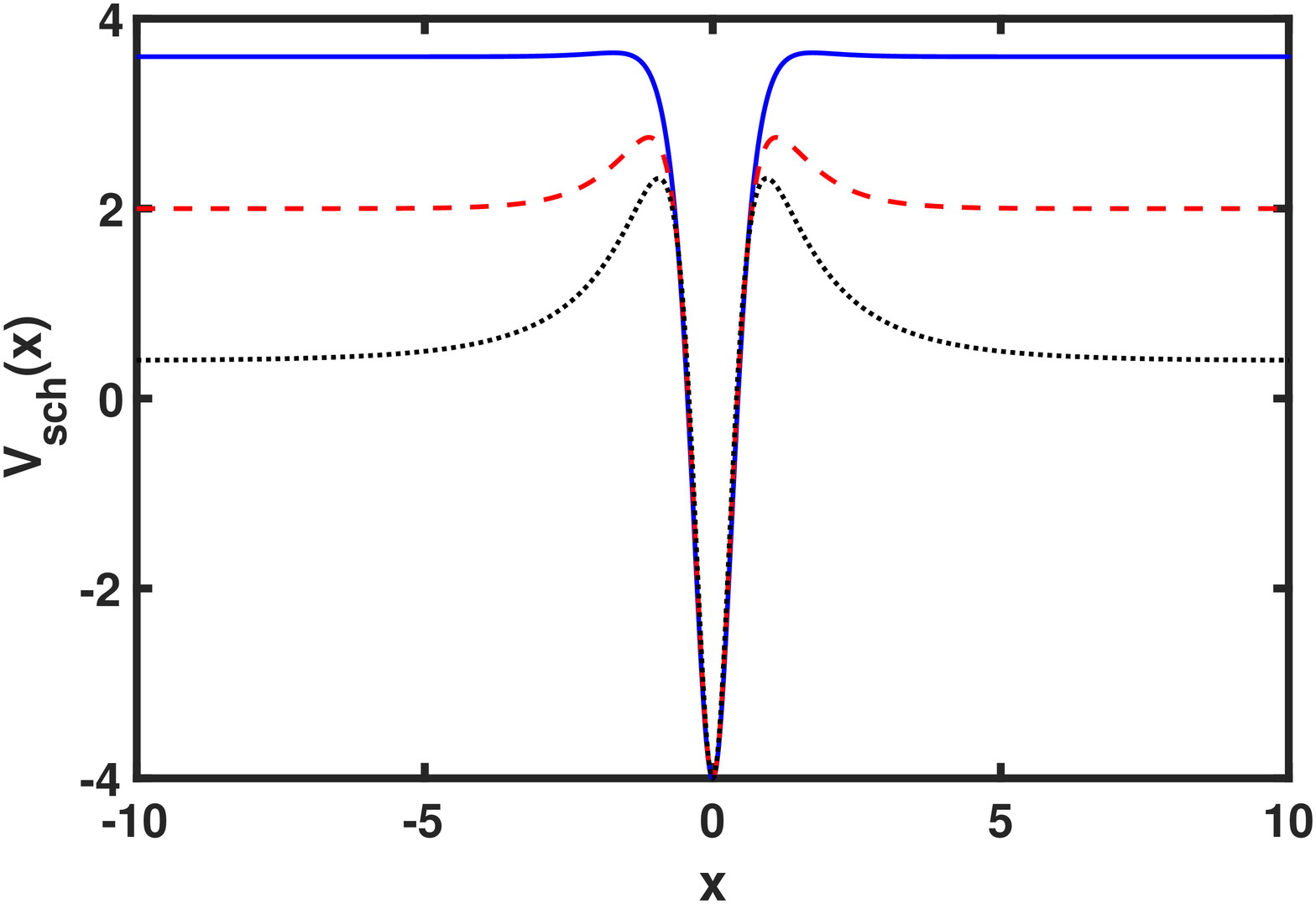}
	\caption{(a) Large kink solutions $\phi^l_K(x)$ and (b) corresponding Schr\"odinger-like potential $V_{sch}(x)$. In both figures we fixed $r=0.1$ (blue solid), $r=0.5$ (red dash), $r=0.9$ (black dot).}
	\label{fig2}
\end{figure}

The large kink solution is given by \cite{baz1}

\be
\phi^l_K (x) = 2\arctan \bigg[ \sqrt{\frac{1+r}{1-r}} \tanh(x \sqrt{1-r}) \bigg],
\ee
and antikink solutions are given by $\phi^l_{\bar K}(x)= \phi^l_K(-x)$. The Fig. \ref{fig2}a shows some plots of large kink profile for several values of $r$. The solutions connect the minima $\pm\phi_v$, with $\phi_v= - \pi + \arccos(r)$.  We note that for large solutions the growth of $r$ contributes to the increasing the asymptotic value of kink. The Fig. \ref{fig2}b depicts plots of Schr\"odinger-like potential $V_{sch}(x)$ for large kink for  several values of $r$. For $r\to0$ we have a central minimum around $x=0$ and asymptotes to $V_{sch}=4$ for $x\to\pm\infty$, corresponding to the sine-Gordon model. This model has a zero-mode followed by a continuum starting at the mass threshold $\omega^2=4$. The increasing of $r$ reduces the asymptotic value of the potential. In particular, for $r\to1$, we have $V_{sch}=0$ for $x\to\pm\infty$. This agrees with ${V_{\phi\phi}}(\phi=\pm\pi)=0$ for $r\to1$, as noted previously. We can see the process of formation of the volcano-shape potential with the growth of parameter $r$. For $0<r<1$ there is only a bound state, the zero mode, followed by a continuum of states at $\omega^2$ corresponding to the asymptotic value of $V_{sch}$. In addition, the volcano-like potential leads to the possibility of resonances. However, since resonances do not completely store energy during the collision process, we have as a consequence  the disappearance of some two-bounces windows \cite{dorey1}.

\begin{figure}
	\includegraphics[{angle=0,width=8cm,height=5cm}]{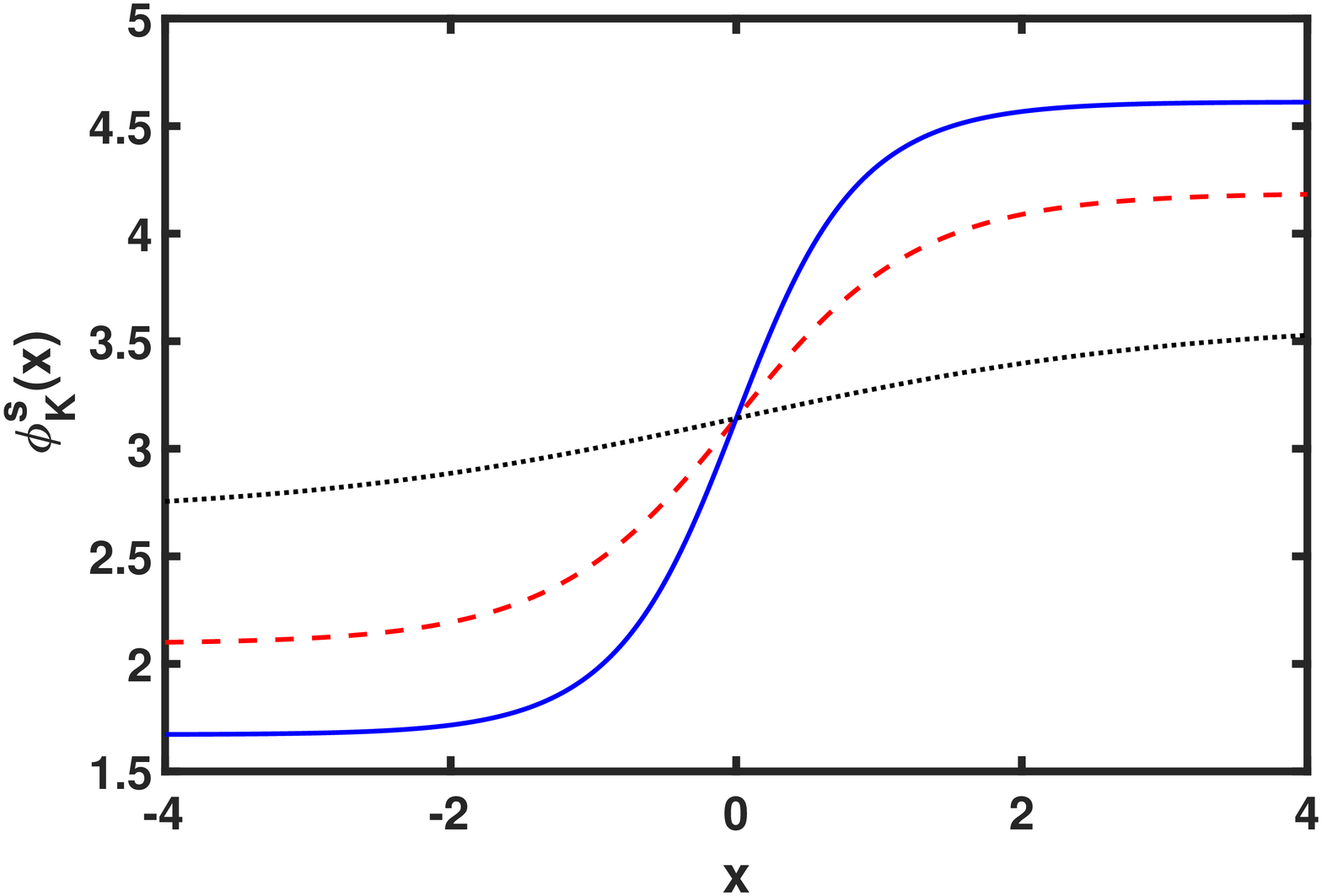}
	\includegraphics[{angle=0,width=8cm,height=5cm}]{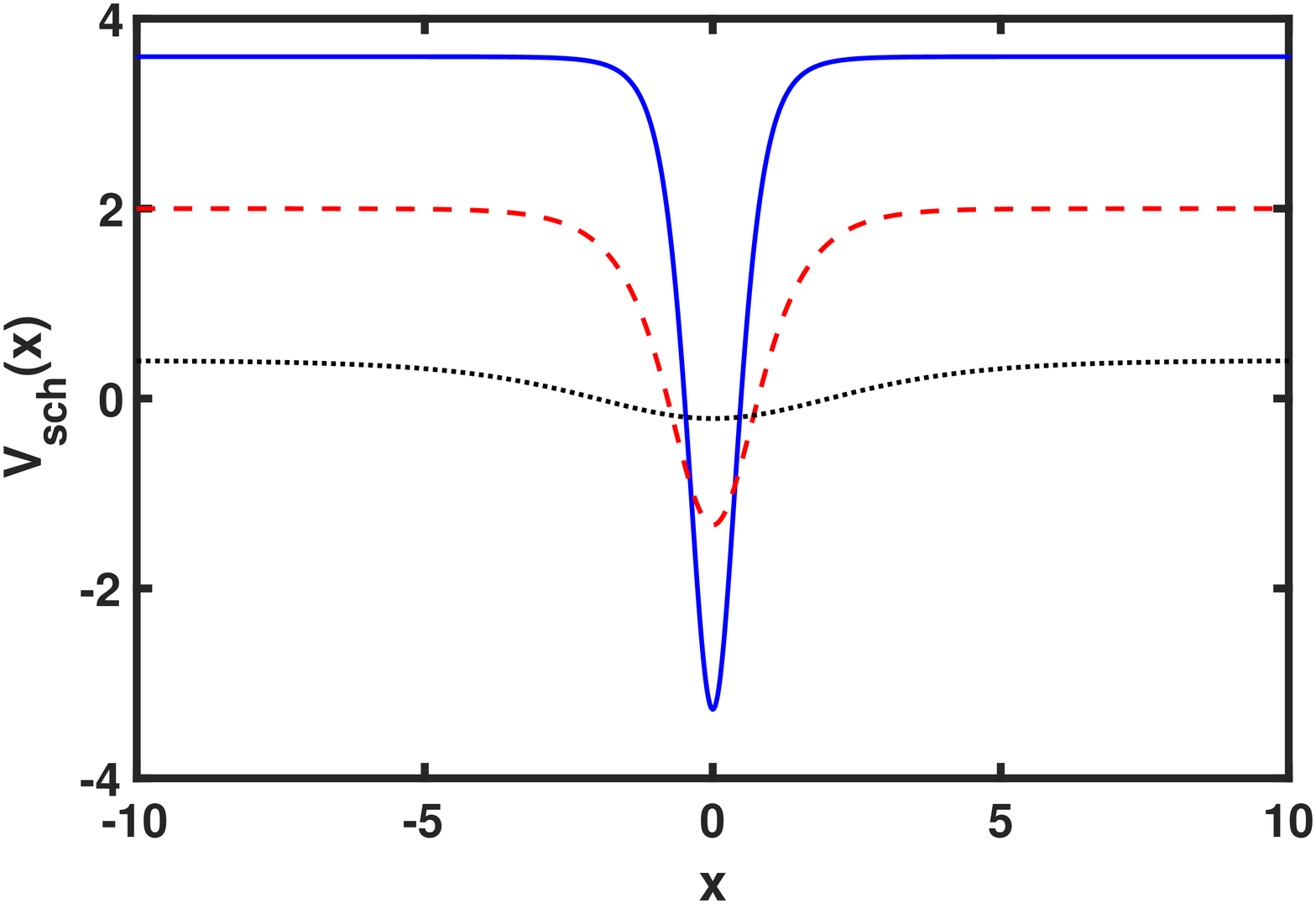}
	\caption{(a) Small kink solutions $\phi^s_K(x)$ and (b) corresponding Schr\"odinger-like potential $V_{sch}(x)$. In the figures we fixed $r=0.1$ (blue solid), $r=0.5$ (red dash), $r=0.9$ (black dot).}
	\label{small}
\end{figure}
The small kink solution is given by \cite{baz1} 

\be
\phi^s_K (x) = \pi+2\arctan \bigg[ \sqrt{\frac{1-r}{1+r}} \tanh(x \sqrt{1-r}) \bigg],
\ee
and antikink solutions are given by $\phi^s_{\bar K}(x)= \phi^s_K(-x)$. The Fig. \ref{small}a shows some plots of small kink profile for several values of $r$. These solutions connect the minima $\phi_v= \pi \pm \arccos(r)$. We note that the increase in the parameter $r$ reduces the value of $\phi_v$.
\begin{figure}

	\includegraphics[{angle=0,width=8cm,height=5cm}]{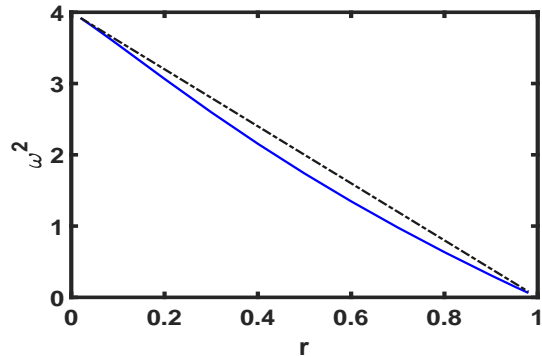}
	\caption{Small kink: squared frequencies $\omega^2$ of vibrational states as a function of the parameter $r$ (blue line). The black dashed line corresponds to the threshold of the continuum modes.}
	\label{estados}
\end{figure}
The Fig. \ref{small}b depicts the Schr\"odinger-like potential $V_{sch}(x)$ for small kink for some values of $r$. We have the same structure of $V_{sch}(x)$ for the antikink $\phi^s_{\bar K}(x)$. We notice that the increase of $r$ reduces the depth of the minimum and decrease the asymptotic maximum of the potential.
 The occurrence of bound states for small kink was investigated numerically for several values of $r$.  There is always a zero mode for all values of $r$. The structure of vibrational mode is summarized in the Fig. \ref{estados}. For $r \gtrsim 0.02$  we have the presence of internal mode, whose squared frequency $\omega^2$ decreases with  $r$. 

\begin{figure}
	\includegraphics[{angle=0,width=8cm,height=5cm}]{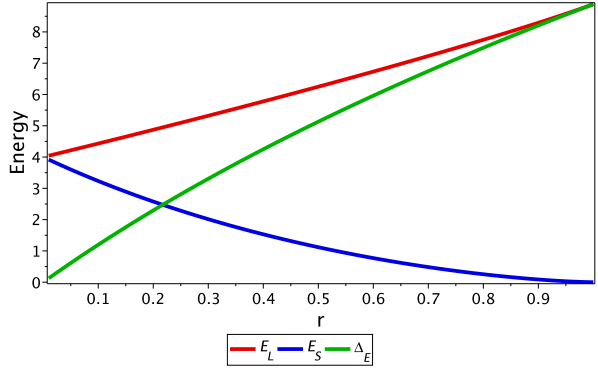}
	\caption{Energy $E_L$ of large kink (red), $E_S$ of small kink (blue) and the difference $\Delta_E=E_L-E_S$ (green) as a function of $r$.}
	\label{E}
\end{figure}
For the same values of $r$, large kinks have larger energies than small kinks. Large kink connecting minima $-\pi +\xi(r)$ and $\pi-\xi(r)$, with energy \cite{baz1}
\be
E_L=4\sqrt{1-r}+4r\frac{\pi-\xi(r)}{\sqrt{1+r}}
\label{EL}
\ee
 and small kink connecting minima $\pi -\xi(r)$ and $\pi+\xi(r)$ with energy  \cite{baz1}
\be
E_S=4\sqrt{1-r}-4r\frac{\xi(r)}{\sqrt{1+r}}.
\label{ES}
\ee
The Fig. \ref{E} shows the plots from $E_L, E_S$ and the difference $E_L-E_S$ as a function of $r$. Note  from the Eqs. (\ref{EL}) and (\ref{ES}) and from the figure that the difference $\Delta_E=E_L-E_S$ has a monotonic increase with $r$. 

For the kink-antikink scattering process we solved the equation of motion with a $4^{th}$ order finite-difference method with a spatial step $\delta x=0.05$. We fixed $x=\pm x_0 = 12$ for the initial symmetric position of the pair. For the time dependence we used a $6^{th}$ order symplectic integrator method, with a time step $\delta t=0.02$. We used the following initial conditions
\begin{eqnarray}
\phi(x,0) & = & \phi_K(x+x_0,v,0) - \phi_K(x-x_0,-v,0) - \phi_v \\
\dot{\phi}(x,0) & = & \dot{\phi}_K(x+x_0,v,0) - \dot{\phi}_K(x-x_0,-v,0),
\end{eqnarray}
where $\phi_K(x,t) = \phi_S(\gamma(x-vt))$ means a boost solution for the particular static kink solution $\phi_S(x)$ (which can be $\phi^l_K (x)$ or $\phi^s_K (x)$), where $\gamma=(1-v^2)^{-1/2}$ and $\phi_v$ is one vacuum of the theory, i.e., a minimum of $V(\phi)$.

In the following we will discuss our main results of scattering for  large and small kinks separately.


\subsection { Large kink scattering }


When  $r=0$ we recover the results of sine-Gordon-like model:  the kink-antikink pass through each other after the first impact.  For all  values of $0<r<1$ our numerical results show the changing of the topological sector after the collision. For $r=0.1$ the outcome is an antikink-kink pair, i.e., the kink-antikink pair approach one each other, collide once and escape for infinity on a new vacuum. The formed antikink-kink pair is accompanied by small oscillations around the inside vacuum. 

\begin{figure}
	\includegraphics[{angle=0,width=7cm}]{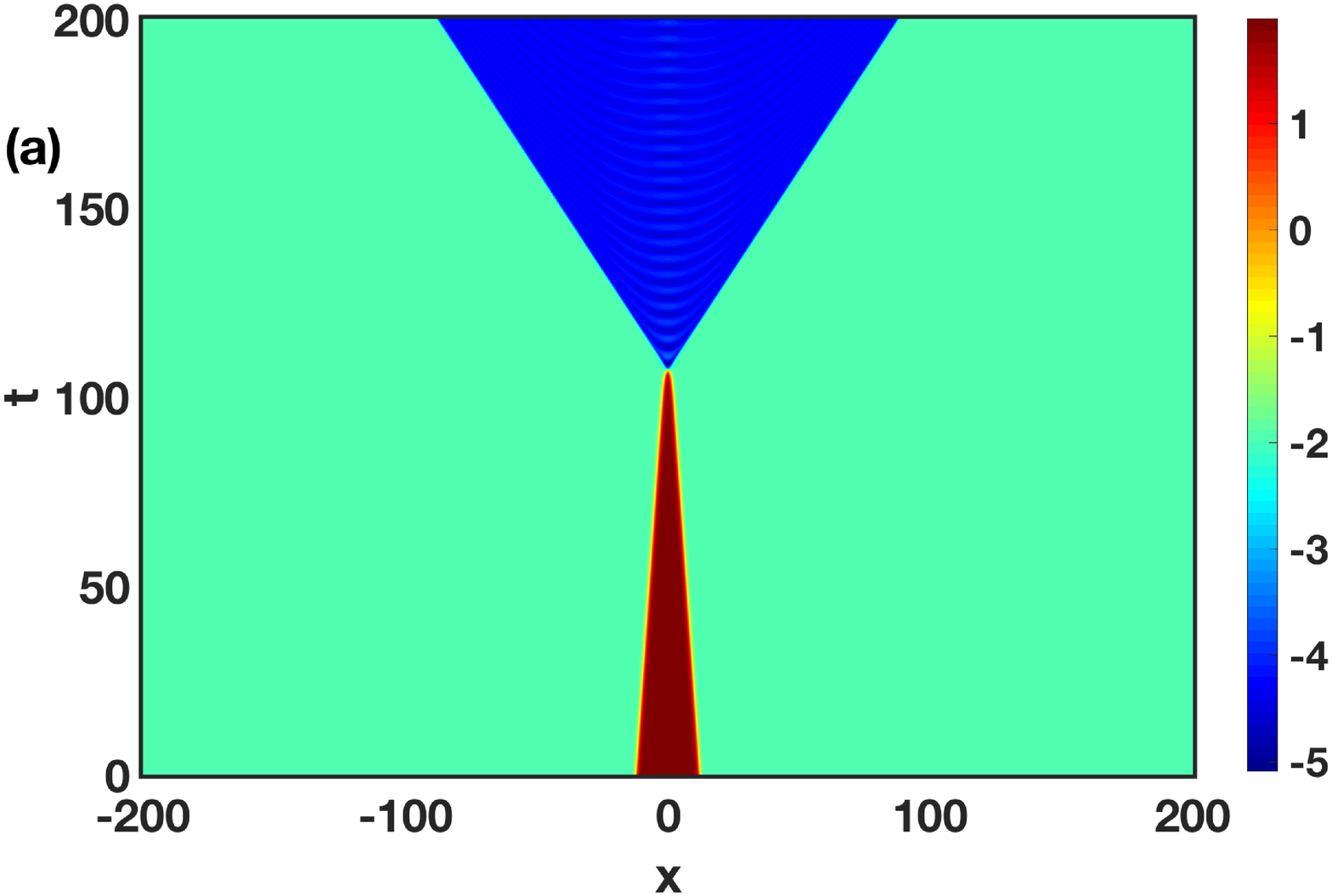}
	\includegraphics[{angle=0,width=7cm}]{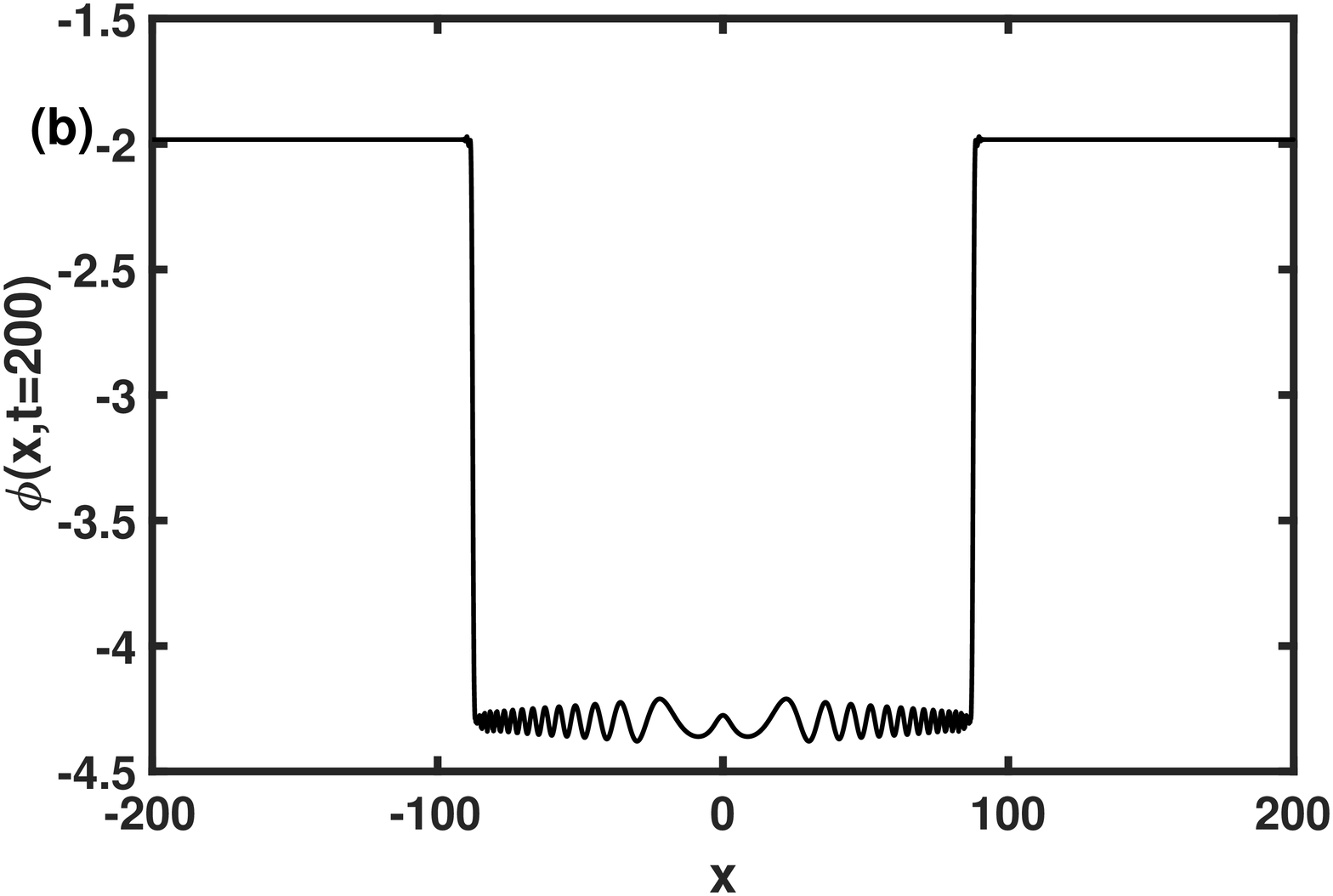}
	\includegraphics[{angle=0,width=7cm}]{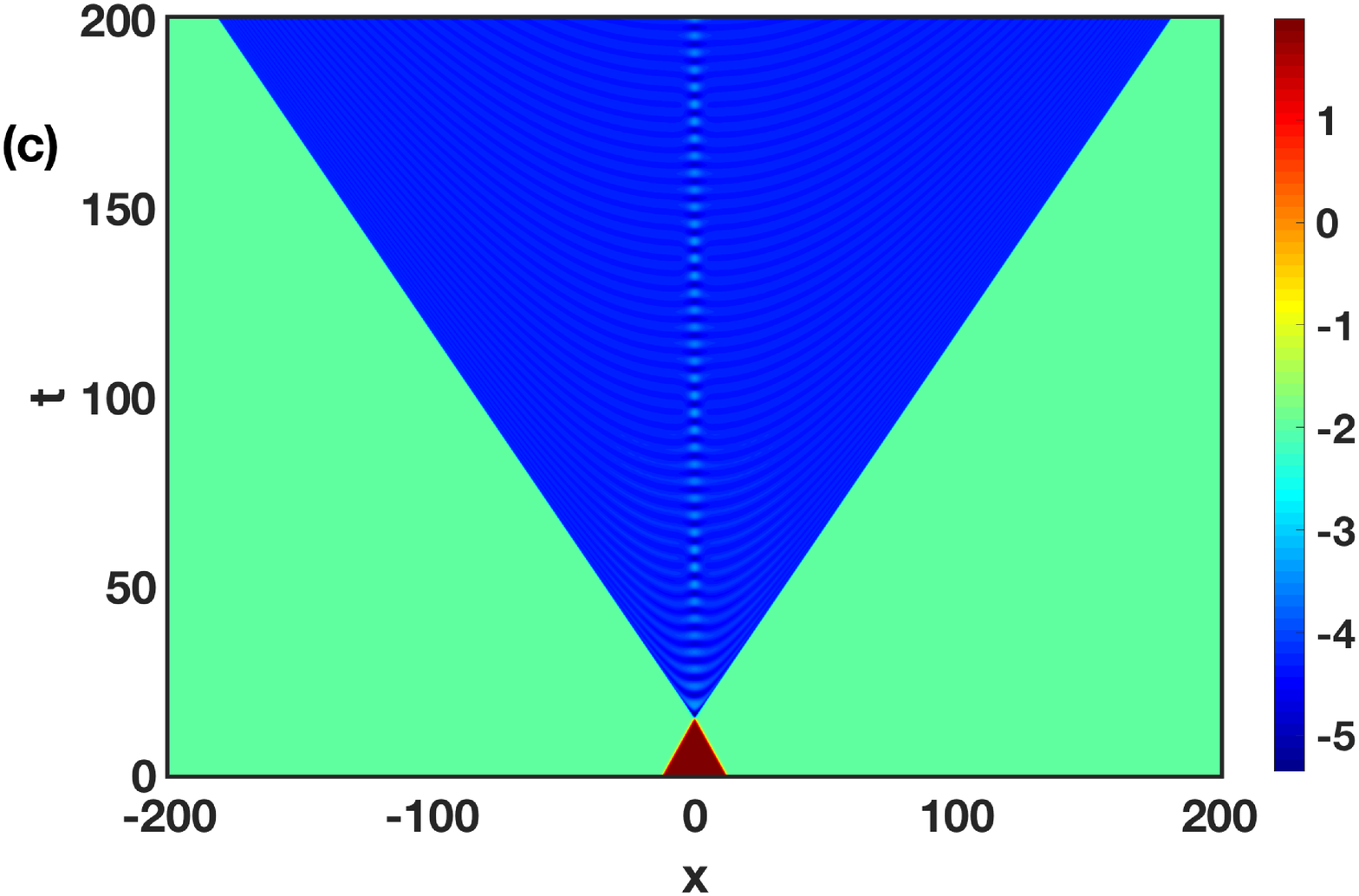}
	\includegraphics[{angle=0,width=7cm}]{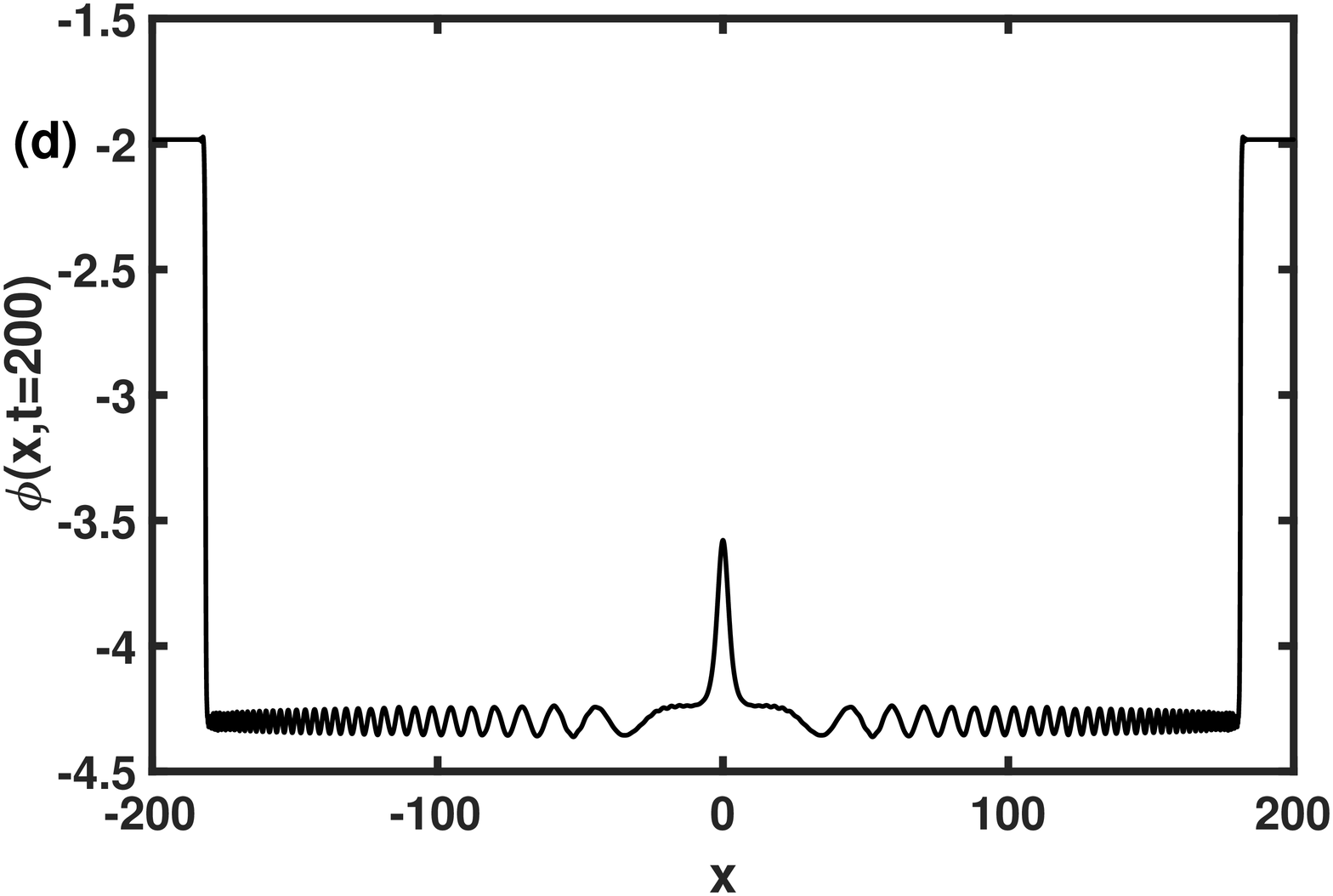}
	\caption{Large kink-antikink collision (left) and image of last simulation time of scalar field as a function of $x$ (right) for a) b) $v=0.1$ and c) d) $v=0.8$. In all figures we fixed $r=0.4$. Note the presence of the confined oscillation for $v=0.8$}
	\label{space}
\end{figure}

\begin{figure}
	\includegraphics[{angle=0,width=7cm}]{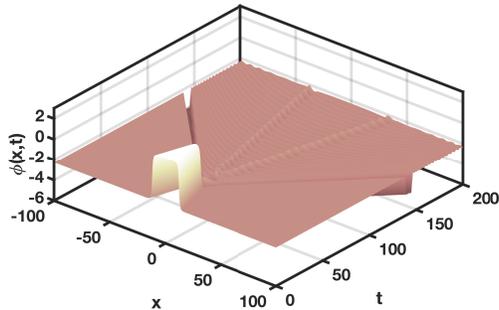}
	\caption{Large kink-antikink collision for $r=0.65$ with $v=0.3$, showing the production of two thin small kink-antikink pairs and two propagating oscillations.}
	\label{phi3d1}
\end{figure}
To better investigate these inside oscillations we considered a large value $r=0.4$. In the Figs. \ref{space}a and \ref{space}c we plot the kink-antikink collisions for $v=0.1$ and $v=0.8$, respectively, and Fig. \ref{space}b and \ref{space}d the image of last simulation time of the scalar field as a function of space $x$ with the same initial velocities.  Note that the velocity of the formed pair is greater than the initial one. The reason is that the initial large kink-antikink changes to small antikink-kink having lower energy (as noted in the Fig. \ref{E}). Then there is enough energy to increase the kinetic energy of the pair. We observe the absence of radiation out of the formed antikink-kink pair. Moreover, the oscillations inside the pair decrease in amplitude and grow in frequency from the center to the border of the antikink-kink pair.  We can interpret that these oscillations are responsible for the formation of antikink-kink pair. This is some sense similar to the Ref. \cite{rosh} on the $\phi^4$ model. In that paper the authors studied the scattering of two identical waves and noted the presence of an oscillon in the collision center as a source of new kink-antikink pairs. Here we see some important differences for  double sine-Gordon model in the sense that i) more antikink-kink pairs can be formed, but only those more external has oscillations with radiation limited to their borders, and ii) it can also be seen the propagation of isolated oscillations.
 One example of this can be seen for $r=0.65$ in the Fig. \ref{phi3d1}. There  we see the production of two thin antikink-kink pairs after the collision. Moreover, we see clearly two oscillatory waves with large harmonicity and correspondingly large lifetime.  
\begin{figure}
	\includegraphics[{angle=0,width=7cm,height=4.5cm}]{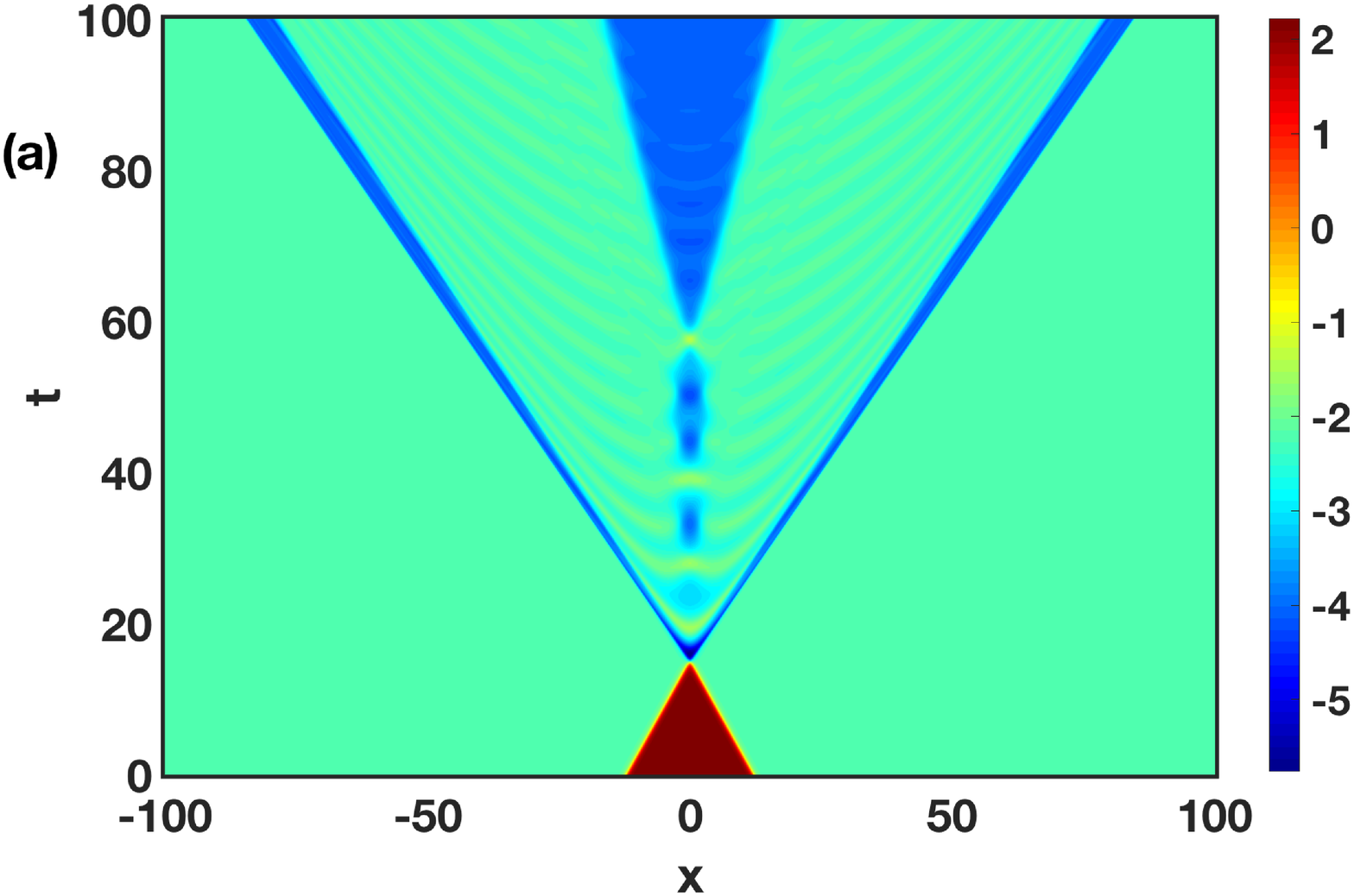}
	\includegraphics[{angle=0,width=7cm,height=4.5cm}]{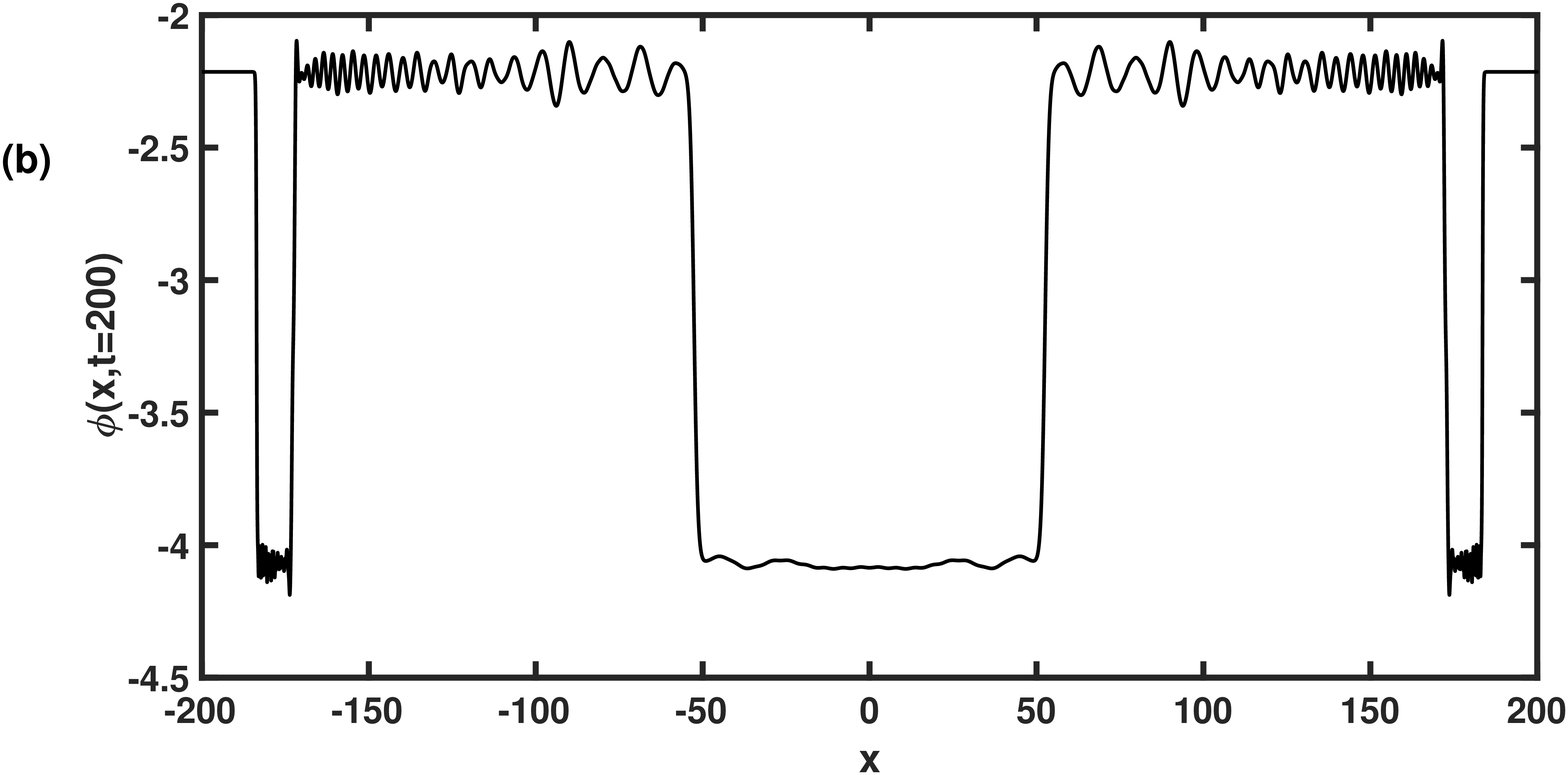}
	\includegraphics[{angle=0,width=7cm}]{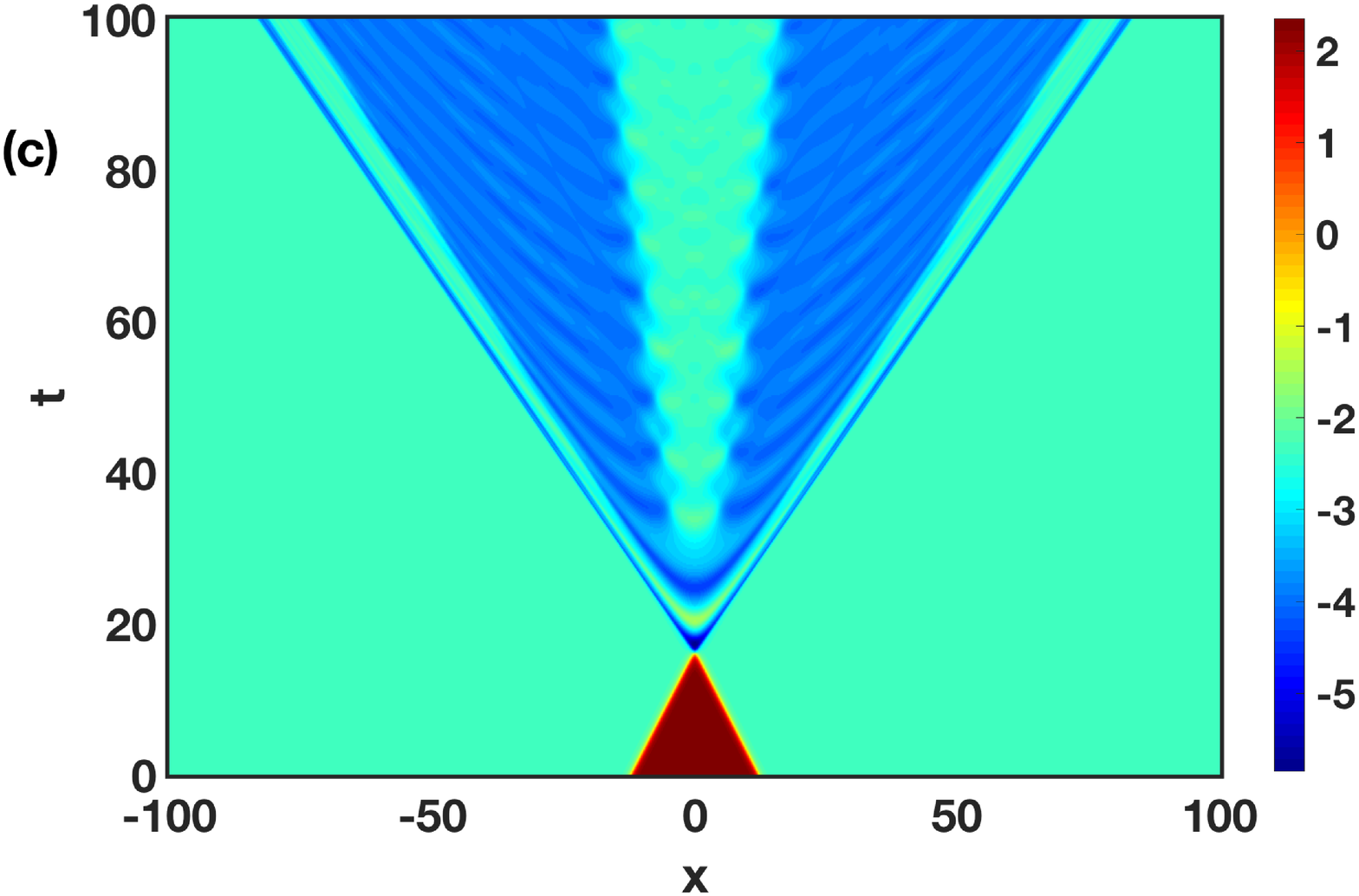}
	\includegraphics[{angle=0,width=7cm,height=4.5cm}]{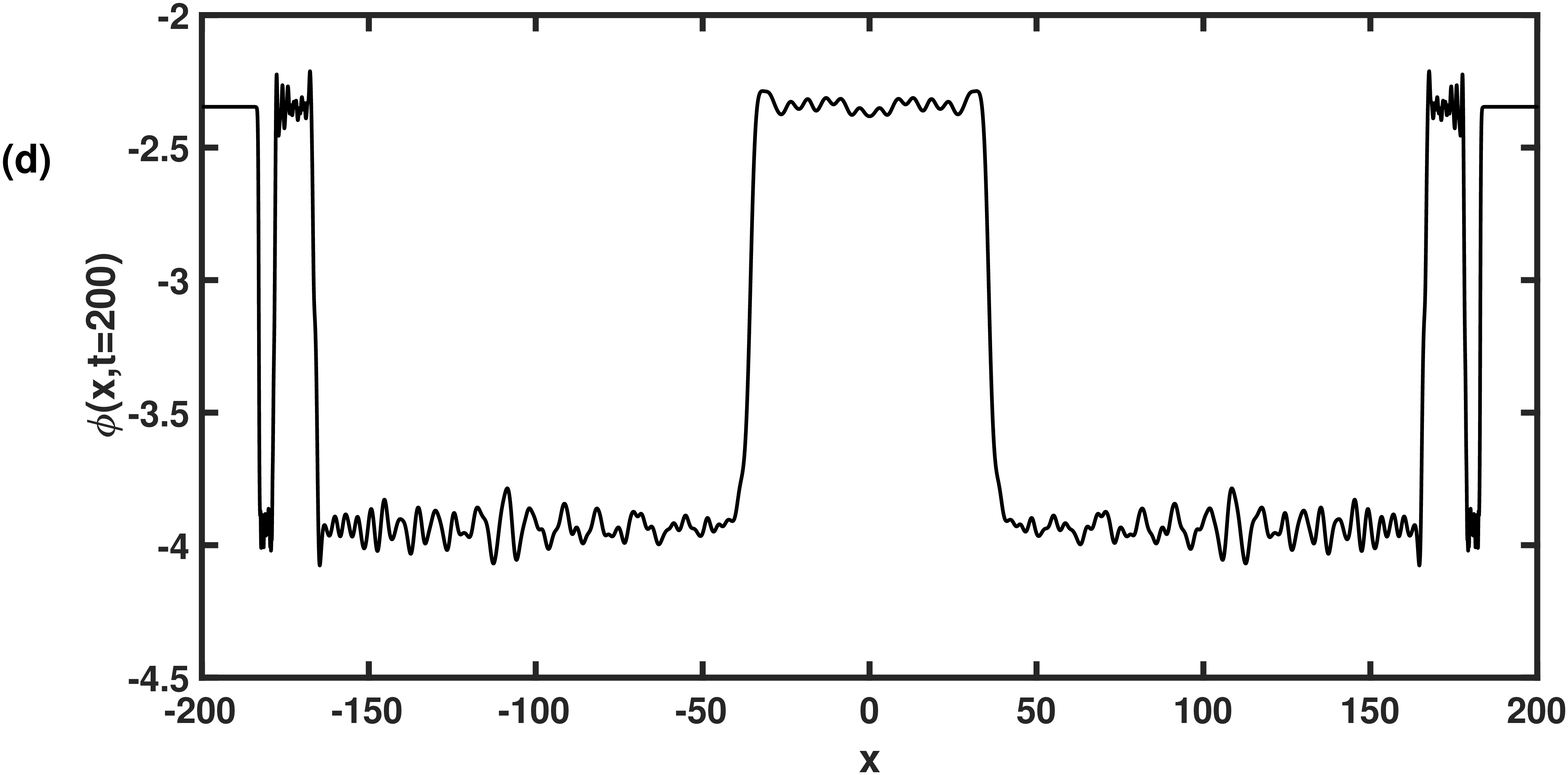}
	\caption{Large kink-antikink collision (left) and image of last simulation time of scalar field as a function of $x$ (right) for a)  b) $r=0.6$, $v=0.8$, with production of three small kink-antikink pairs and c) d) $r=0.7$, $v=0.74$, with production of four small kink-antikink pairs.}
	\label{space2}
\end{figure}

The  production of new antikink-kink pairs is more sensitive to the parameter $r$ than to the initial velocity. For example, in the Fig. \ref{space2}a-b for $r=0.6$ we can see the formation of three antikink-kink pairs after one collision: two thin and one thick pair. The increase of $r$ favors the production of more antikink-kink pairs, as can be seen in the Fig. \ref{space2}c-d for $r=0.7$, where four antikink-kink pairs are formed. We also note that, for fixed $r$, the more external antikink-kink pairs are the thinnest. Moreover, the increasing of $r$ results in the reduction of the thickness of the more external antikink-kink pair (compare the Figs. \ref{space2}a and \ref{space2}c). 

\begin{figure}
	\includegraphics[{angle=0,width=16cm,height=6cm}]{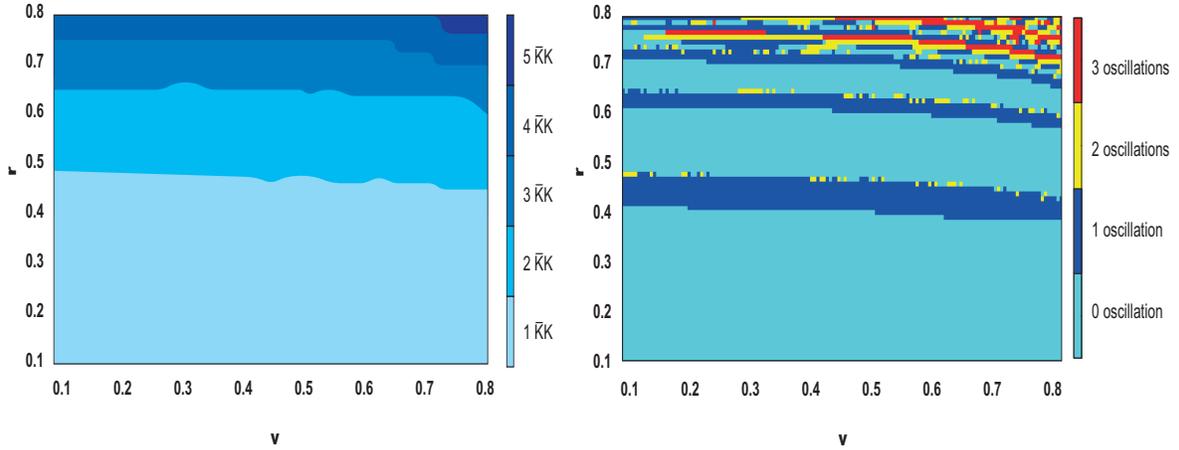}
	\caption{Output of the large kink-antikink collision process: a) (left) number of formed small antikink-kink pairs, b) (right) number of solitary oscillations.}
	\label{mosaico}
\end{figure}

The number of formed antikink-kink pairs at the output of the collision as a function of $r$ and $v$ is presented in Fig. \ref{mosaico}a. The figure shows that small values of $r$ are related to the formation of only one antikink-kink pair. There is a transition region $r\sim 0.5$ such that for $0.5 \lesssim r \lesssim 0.65$ we have two antikink-kink pairs. Increasing more the parameter $r$ we see two more transition regions, $r\sim 0.65$ and $r\sim 0.75$ corresponding respectively to the formation of three (for $0.65 \lesssim r \lesssim 0.75$) and four (for $0.75 \lesssim r \lesssim 0.8$) antikink-kink pairs. That is,  larger values for $ r $ favors the occurrence of a larger number of pairs.

In the Fig. \ref{mosaico}b we can see the number of solitary oscillatory waves due to the collision process. Note that small values of $r$ are not related to the observation of these oscillations. Increasing $r$ there is the possibility of occurrence of oscillations, but only in some intervals of $r$ that  follows closely the transition regions described in the Fig.  \ref{mosaico}a for the number of antikink-kink pairs. For example, the Fig.  \ref{mosaico}b shows that, for $0.4 \lesssim r \lesssim 0.5$ we can see the formation of one oscillation centered at $x=0$, with regions with two oscillations for $r\sim 0.5$. This value of $r$ is roughly in the transition region between the formation of one and two antikink-kink pairs (compare with the Fig.  \ref{mosaico}a). The same applies for $0.62 \lesssim r \lesssim 0.65$, with two oscillations for some velocities at $r\sim 0.65$. This value of $r$ is roughly in the transition region between the formation of three  and four antikink-kink pairs (compare with the Fig.  \ref{mosaico}a). That is, it seems to have a connection between oscillation formation and the increasing in the number of formed antikink-kink pairs. 
\begin{figure}
	\includegraphics[{angle=0,width=8cm}]{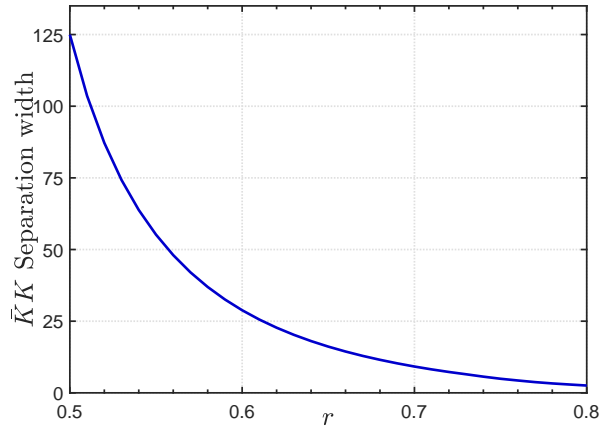}
	\caption{Large kink-antikink collision: average width value of the more external small antikink-kink pair. }
	\label{largura}
\end{figure}

Comparing the Figs. \ref{space} and \ref{space2}, we see that when we have only one produced antikink-kink pair (Fig. \ref{space}), it does not radiate, whereas when we have three or four formed antikink-kink pairs (Fig. \ref{space2}), only the more external pair does not  radiate. In the former case, this happens for a thick pair, whereas in the latter, for a thin pair. That is, the thickness of the antikink-kink pair is not a determinant aspect for the absence of emitted radiation. However, since thin antikink-kink pairs that do not radiate have some resemblance with propagating particles, their observation is of more interest. Due to this we investigated some features regarding the width and the field radiation at the midpoint of the thinnest and more external formed antikink-kink pair. The Fig. \ref{largura}a shows that the average value of the width of the more external antikink-kink pair decreases with the increment of $r$. Compare for instance the Figs. \ref{space2}a and \ref{space2}c, corresponding  respectively to $r=0.6$ and $r=0.74$. 
\begin{figure}
\includegraphics[{angle=0,width=10cm}]{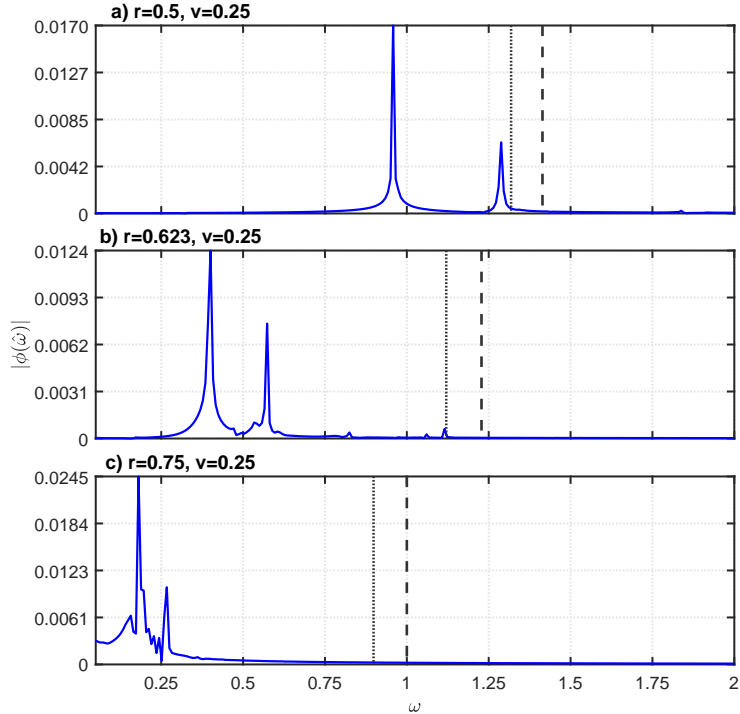} 
	\caption{Large kink-antikink collision: Fourier transform at the center of the thinnest small antikink-kink pair for $100<t<800$. For all points we fixed $v=0.25$. In each figure, the dashed line corresponds to the threshold of the continuum modes.  The dotted  line corresponds to the vibrational mode.}
	\label{fft-pair}
\end{figure}

\begin{figure}
	\includegraphics[{angle=0,width=8cm}]{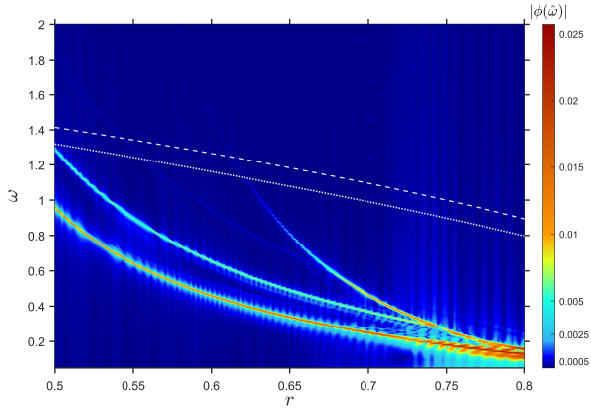}
	\caption{ Large kink-antikink collision: Fourier transform at the center of the thinnest antikink-kink pair. For all points we fixed $v=0.25$. The white dashed line corresponds to frequencies in continuum mode. The white dotted  line corresponds to the vibrational mode. }
	\label{freq}
\end{figure}
The Figs.  \ref{fft-pair}a-c depicts the Fourier transform of the thinnest antikink-kink pair for a specific initial velocity. The figures show that the frequencies are below the continuum, meaning that the pair is not able to radiate. For low values of $r$ we have two frequencies (Fig. \ref{fft-pair}a), and we noticed a decline in these frequencies with the growth of $r$. A third frequency appears for $ r\gtrsim 0.623$ (Fig. \ref{fft-pair}b). This region coincides with an increase in the number of formed kink-antikink pairs. This behavior is roughly independent of $v$. We observed that changes in the initial velocity modify the amplitude of the field, not the frequencies.  The general behavior of the frequencies with $r$ is depicted in the Fig. \ref{freq}. Note that the frequencies decrease with $r$, and are always below the mass threshold. Moreover, the frequencies are usually below that of the vibrational mode (dotted line of the Fig. \ref{freq}), except for specific regions $r\sim0.5$ and $r\sim 0.64$, close respectively, to the transition region of formation of one oscillation and two oscillations.

\begin{figure}
\includegraphics[{angle=0,width=10cm}]{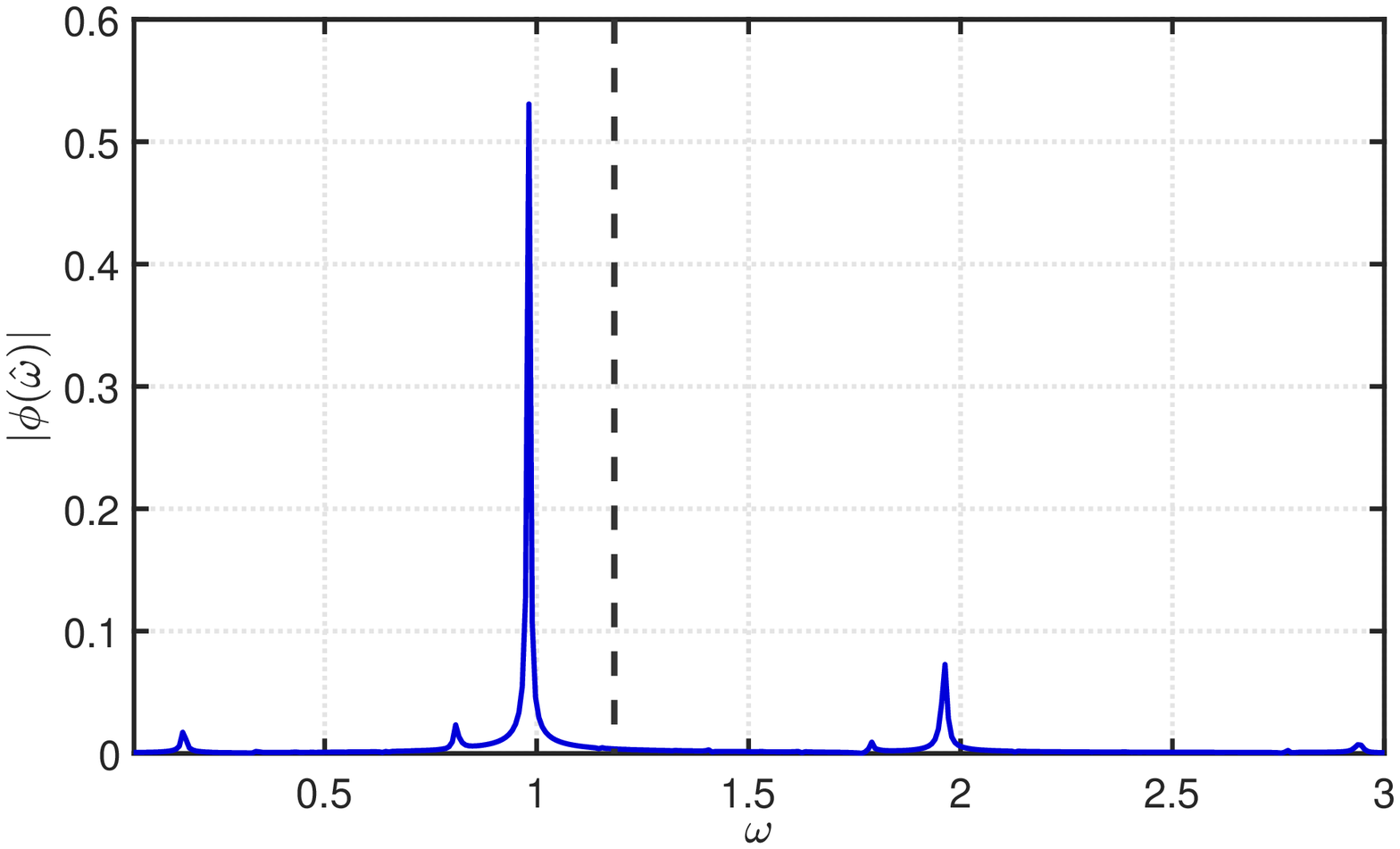}
\includegraphics[{angle=0,width=10cm}]{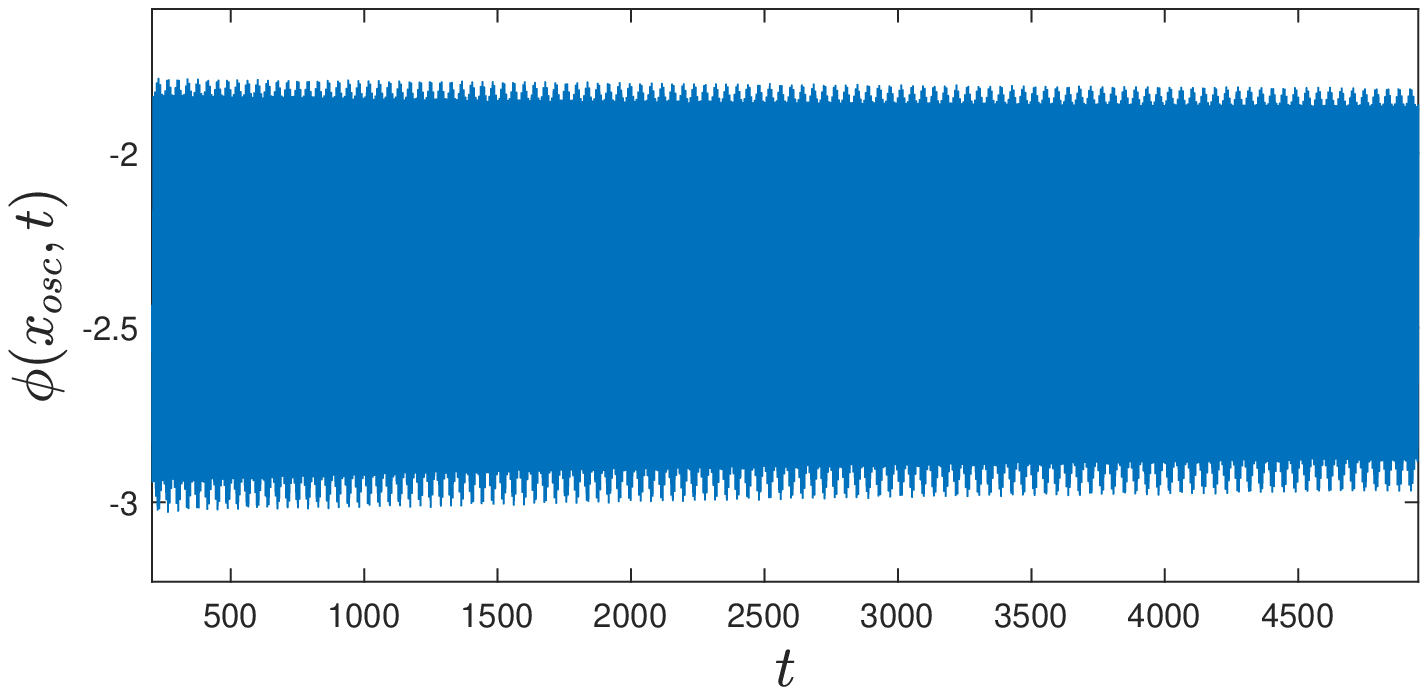}
	\caption{Large kink-antikink collision: a) Fourier transform of one propagating oscillation for $100<t<800$ shown in the Fig. \ref{phi3d1}.  The dashed line corresponds to the threshold of the continuum modes. b) Oscillations of the Fig. \ref{phi3d1} up to  $t=5000$. }
\label{fft-oscill}
\end{figure}
We also analyzed the solitary oscillations. The Fig. \ref{fft-oscill}a shows the Fourier transform of one of the two oscillations shown in the Fig. \ref{phi3d1}. We note from the figure that the main frequency is in the discrete spectrum. Also, it appears one small peak in the continuum.
The amplitude of the oscillations decays linearly, as shown for instance in the Fig. \ref{fft-oscill}b. This means a larger rate than the radiation  of an oscillating $\phi^4$ kink, which follows the Manton-Merabet pattern \cite{mm}.


\subsection {Small kink scattering}

\begin{figure}
	\includegraphics[{angle=0,width=8cm,height=5cm}]{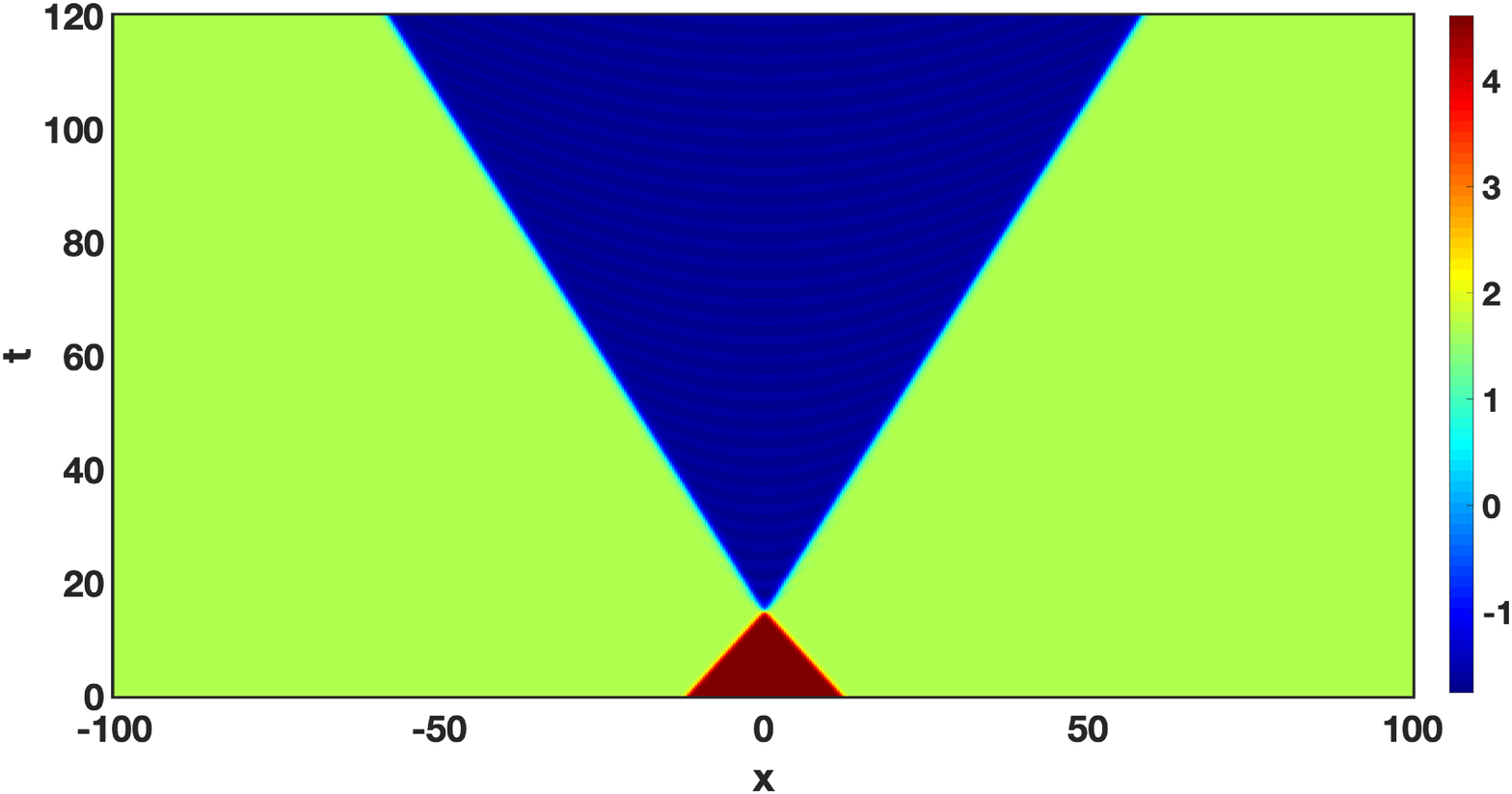}
\includegraphics[{angle=0,width=8cm,height=5cm}]{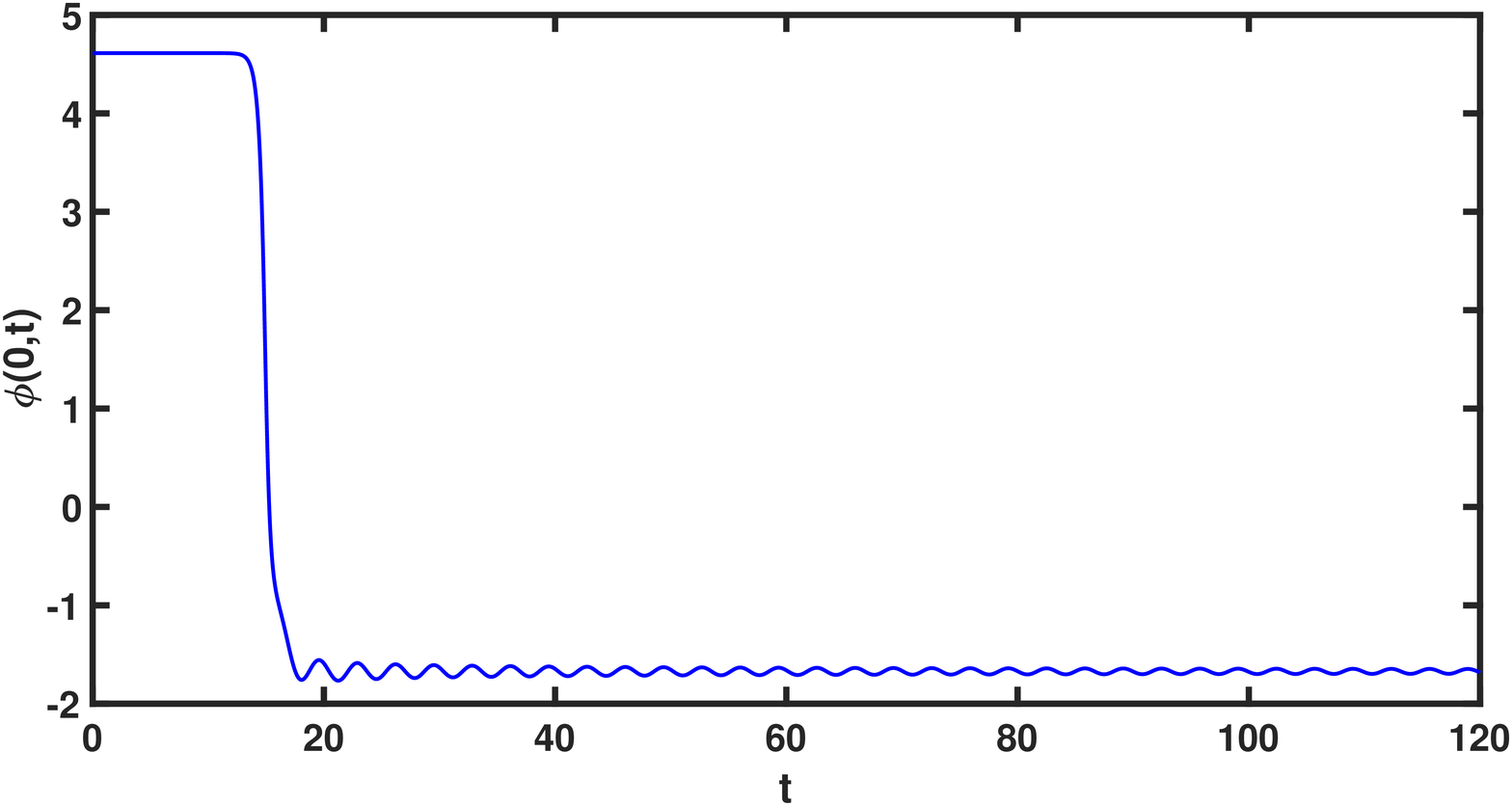}
	\caption{Small kink-antikink collision, showing a) the changing of the topological sector and the production of a large antikink-kink pair. b) The scalar field profile at $x=0$ showing the small oscillations after the collision. In both figures, $r=0.1$ and $v=0.8$. }
	\label{small-large}
\end{figure}

\begin{figure}
	\includegraphics[{angle=0,width=8cm,height=5cm}]{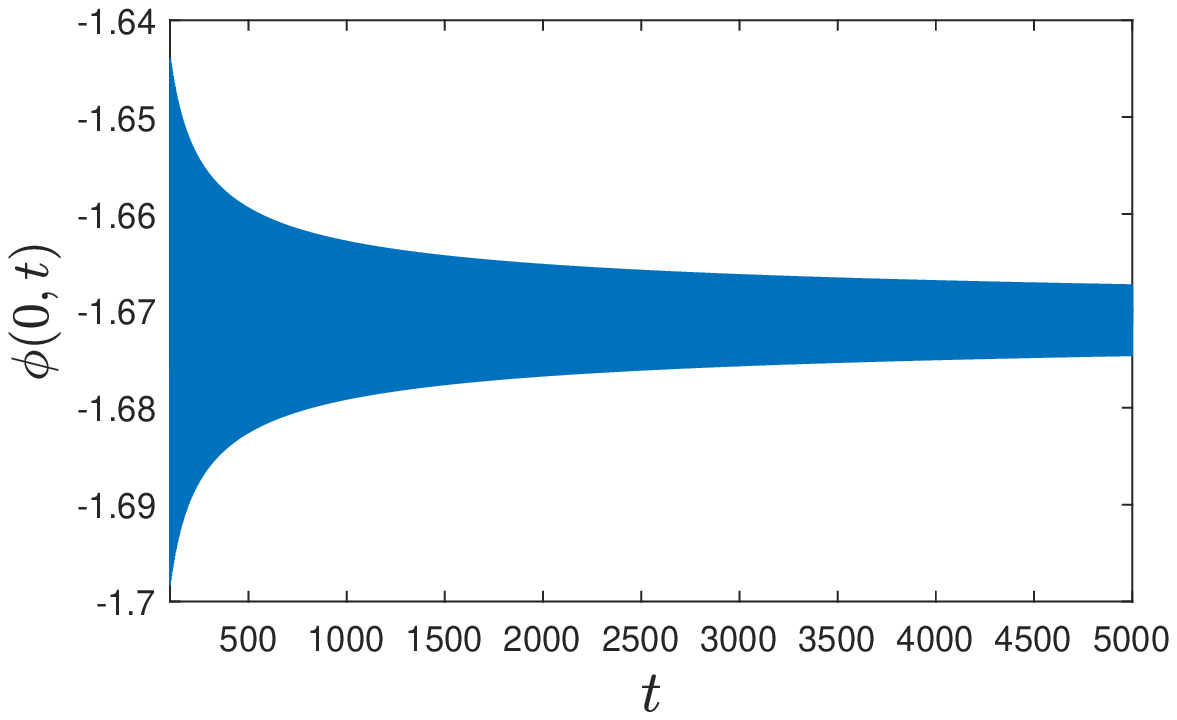}
\includegraphics[{angle=0,width=8cm,height=5cm}]{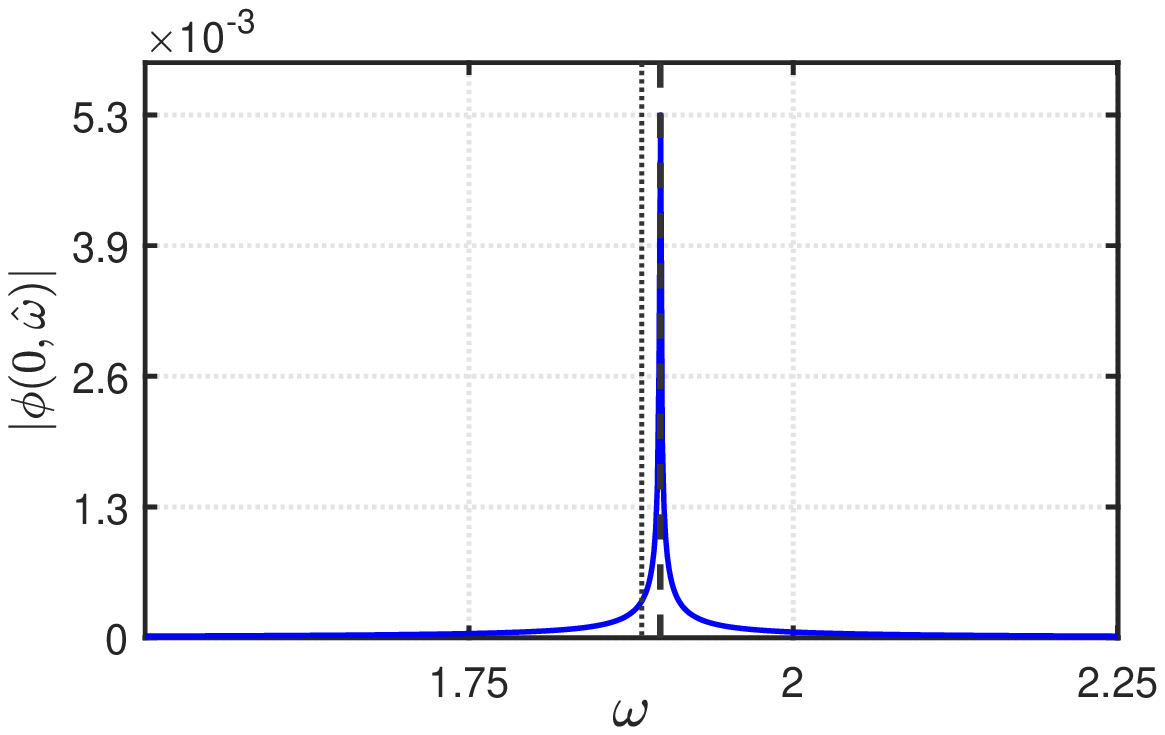}
	\caption{Small kink-antikink collision:  a) Oscillations of the Fig. \ref{small-large}b up to  $t=5000$; b) Fourier transform the oscillations for $200<t<5000$.  The dashed line corresponds to the threshold of the continuum modes.  In both figures, $r=0.1$ and $v=0.8$. }
	\label{small-large-fft}
\end{figure}

The small kink-antikink scattering for small values of $r$ and $v>v_*$ can produce large antikink-kink, as depicted in the Fig.  \ref{small-large}a. Note from the figure that the velocity of the produced large antikink-kink pair is lower than that of the incident small kink-antikink pair. This agrees with the large energy of the large kink in comparison with the small kink (see the Fig. \ref{E}). Inside the formed large antikink-kink pair one can see emitted radiation. The Fig.  \ref{small-large}b shows the value of $\phi(0,t)$ as a function of $t$. One can see the decay of the amplitude of the field after it changes from the topological sector. The Fig. \ref{small-large-fft}a shows $\phi(0,t)$ for a large interval of time.  
The Fourier transform depicted in the Fig. \ref{small-large-fft}b shows that these oscillations are in the continuum region of the large kink spectrum. The amplitude $A$ of the field at $x=0$ follows the decay expression $|dA/dt|\propto A^3$ from Manton-Merabet \cite{mm}.

\begin{figure}
	\includegraphics[{angle=0,width=8cm,height=5cm}]{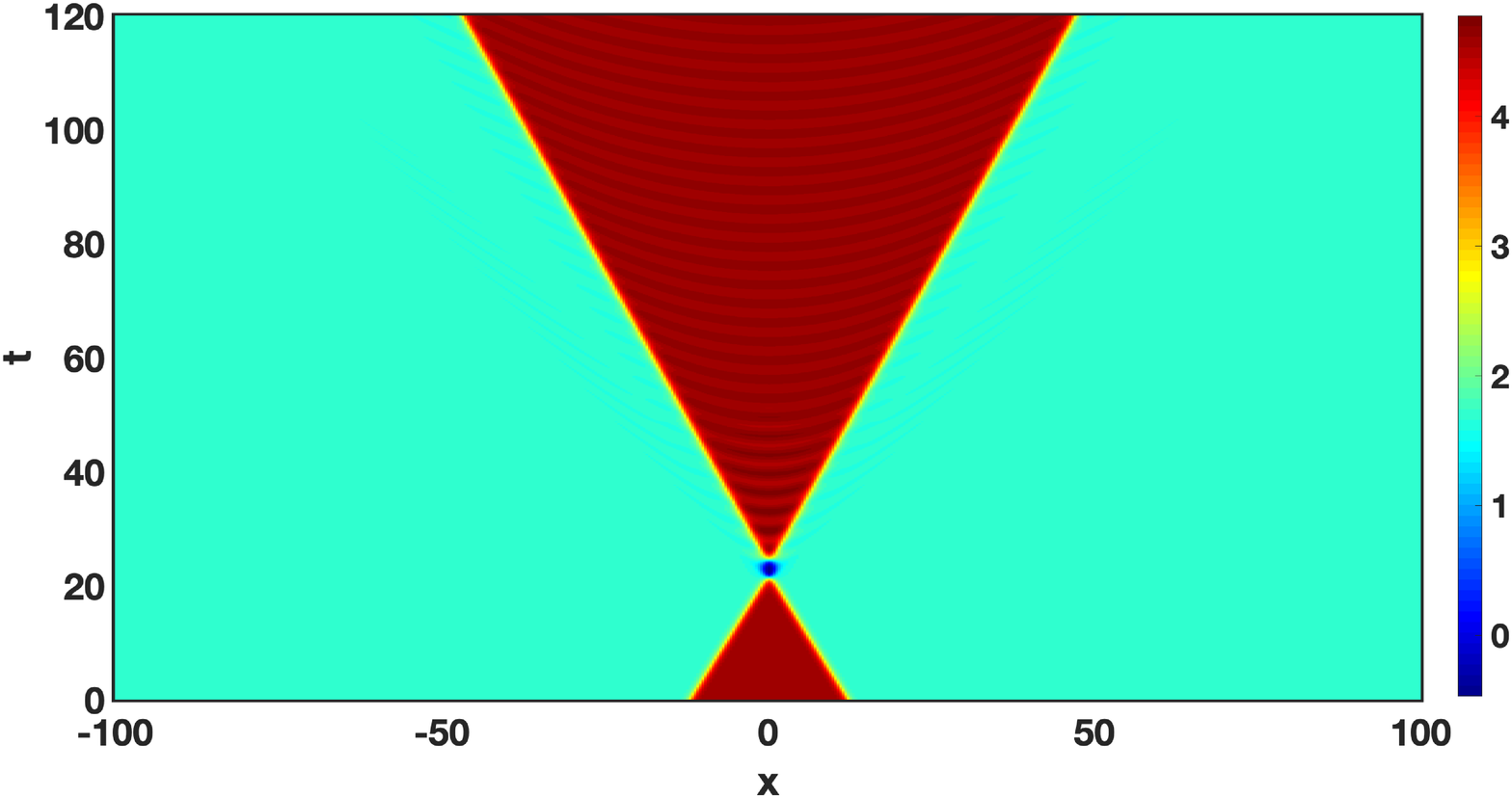}
\includegraphics[{angle=0,width=8cm,height=5cm}]{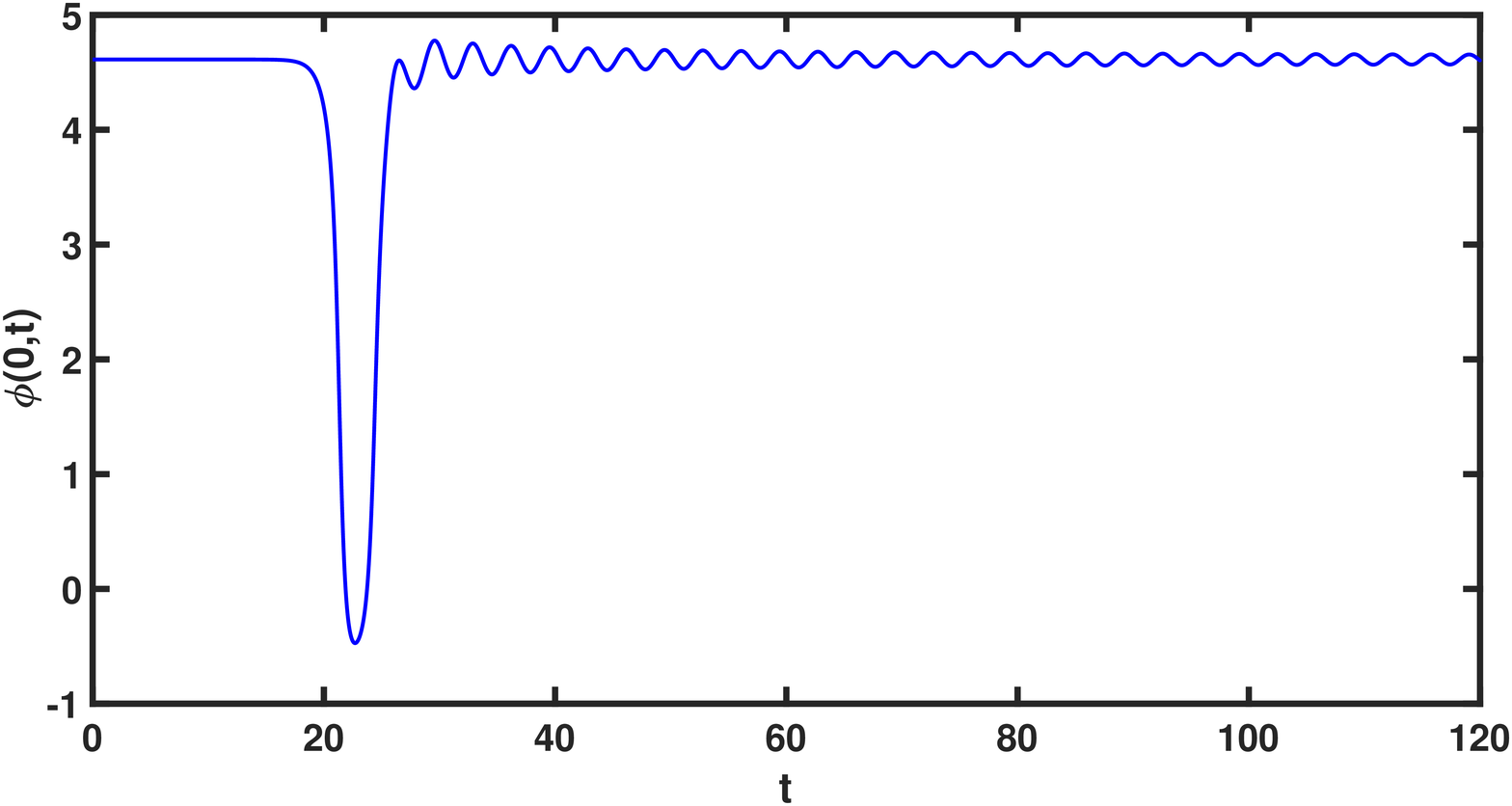}
	\caption{Small kink-antikink collision, showing one bounce: $r=0.1$, $v=0.55$.}
	\label{small-small}
\end{figure}

\begin{figure}
	\includegraphics[{angle=0,width=8cm,height=5cm}]{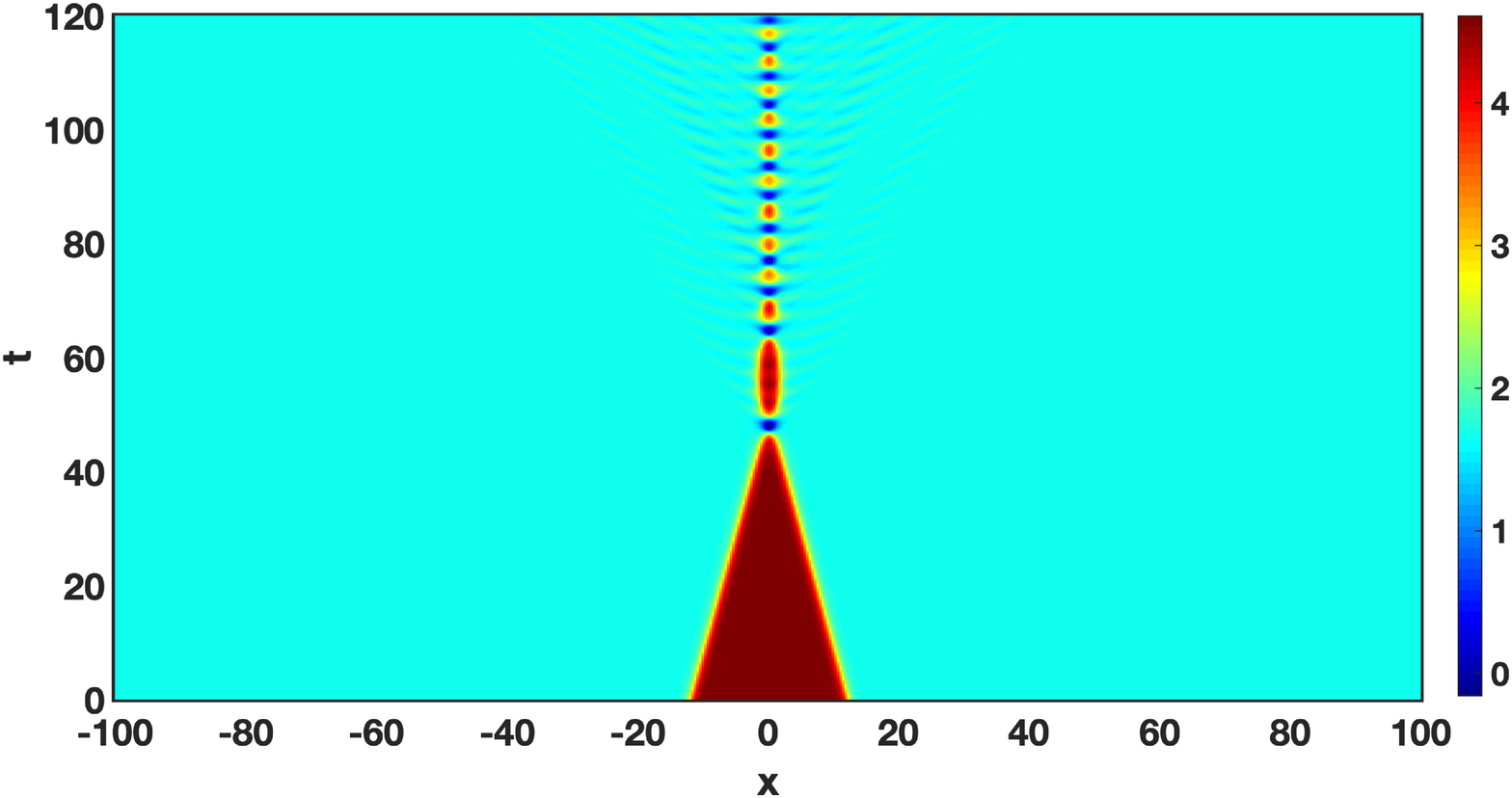}
\includegraphics[{angle=0,width=8cm,height=5cm}]{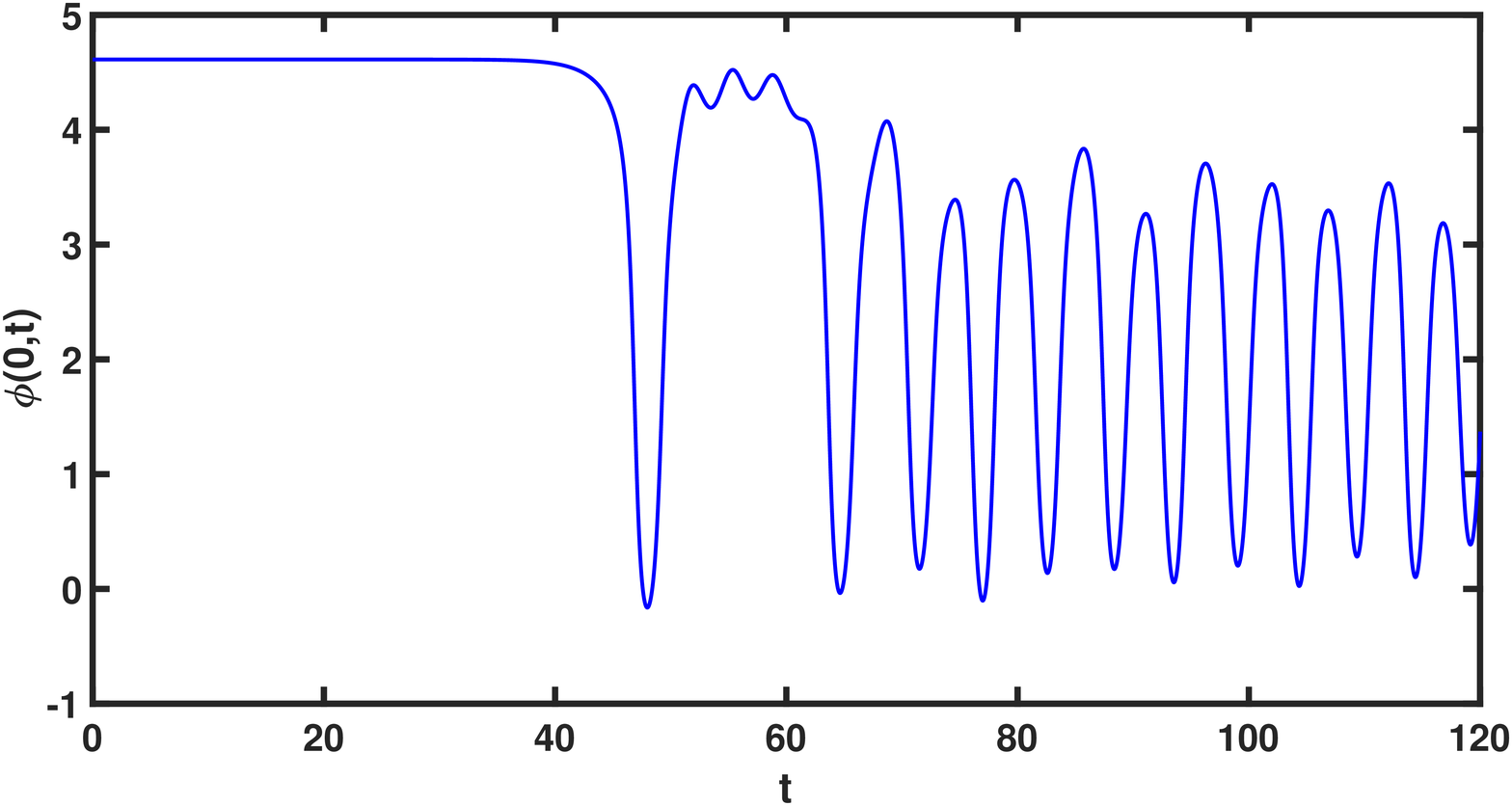}
	\caption{Small kink-antikink collision, producing a bion state: $r=0.1$, $v=0.24$.}
	\label{small-bion}
\end{figure}

\begin{figure}
	\includegraphics[{angle=0,width=8cm,height=5cm}]{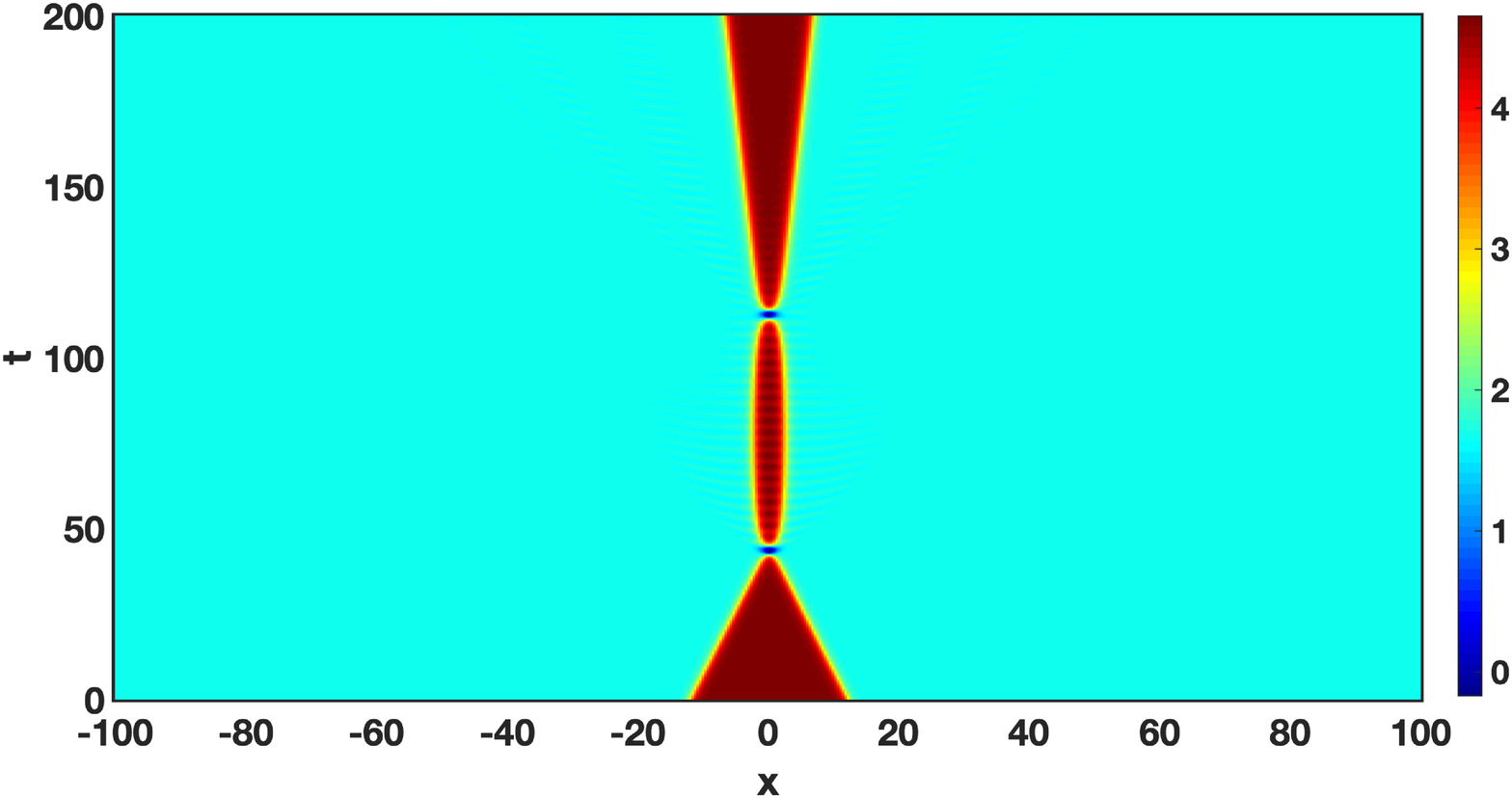}
\includegraphics[{angle=0,width=8cm,height=5cm}]{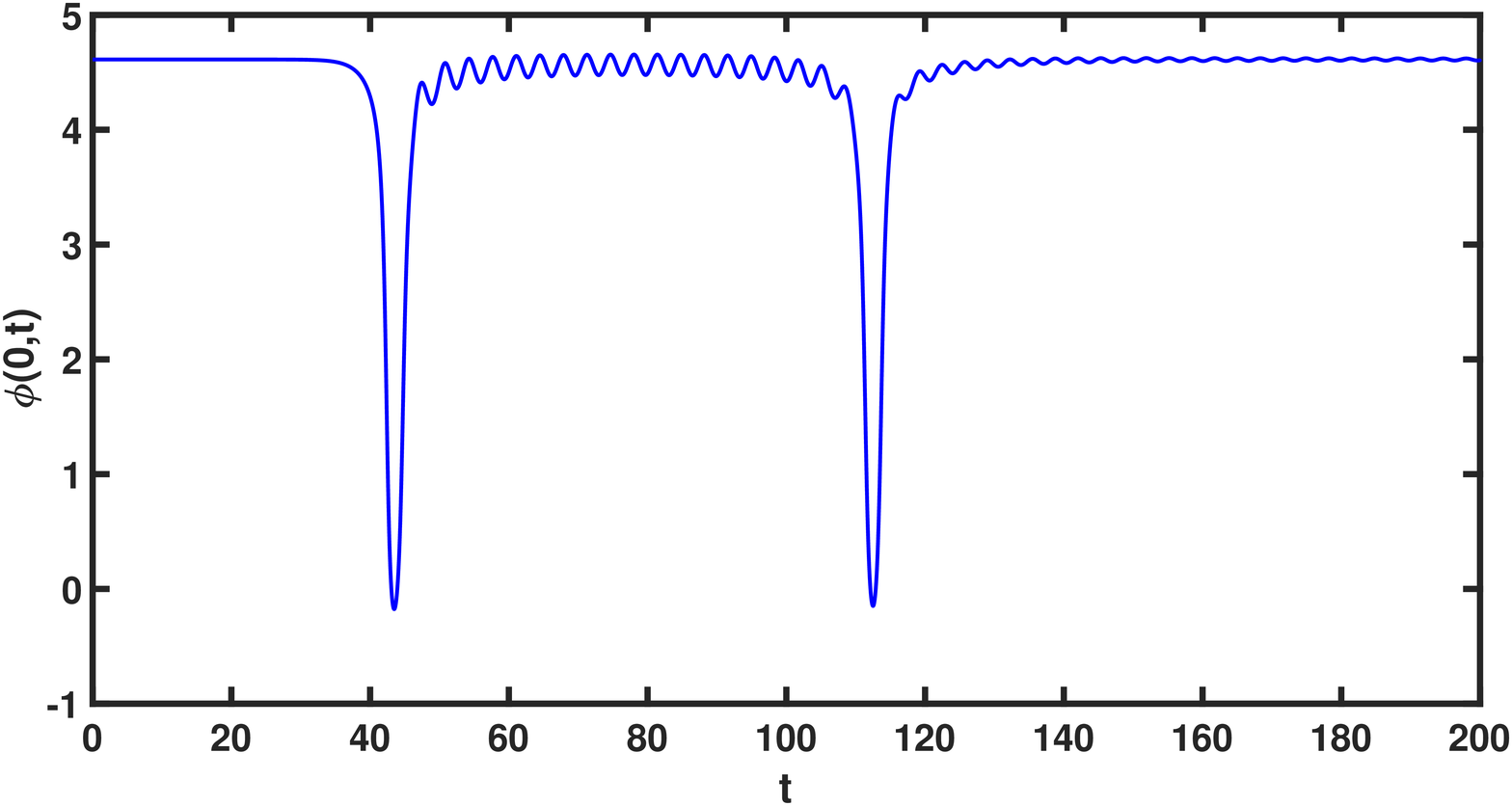}
	\caption{Small kink-antikink collision, showing two bounces: $r=0.1$, $v=0.267$.}
	\label{small-2bounce}
\end{figure}

\begin{figure}
	\includegraphics[{angle=0,width=8cm,height=5cm}]{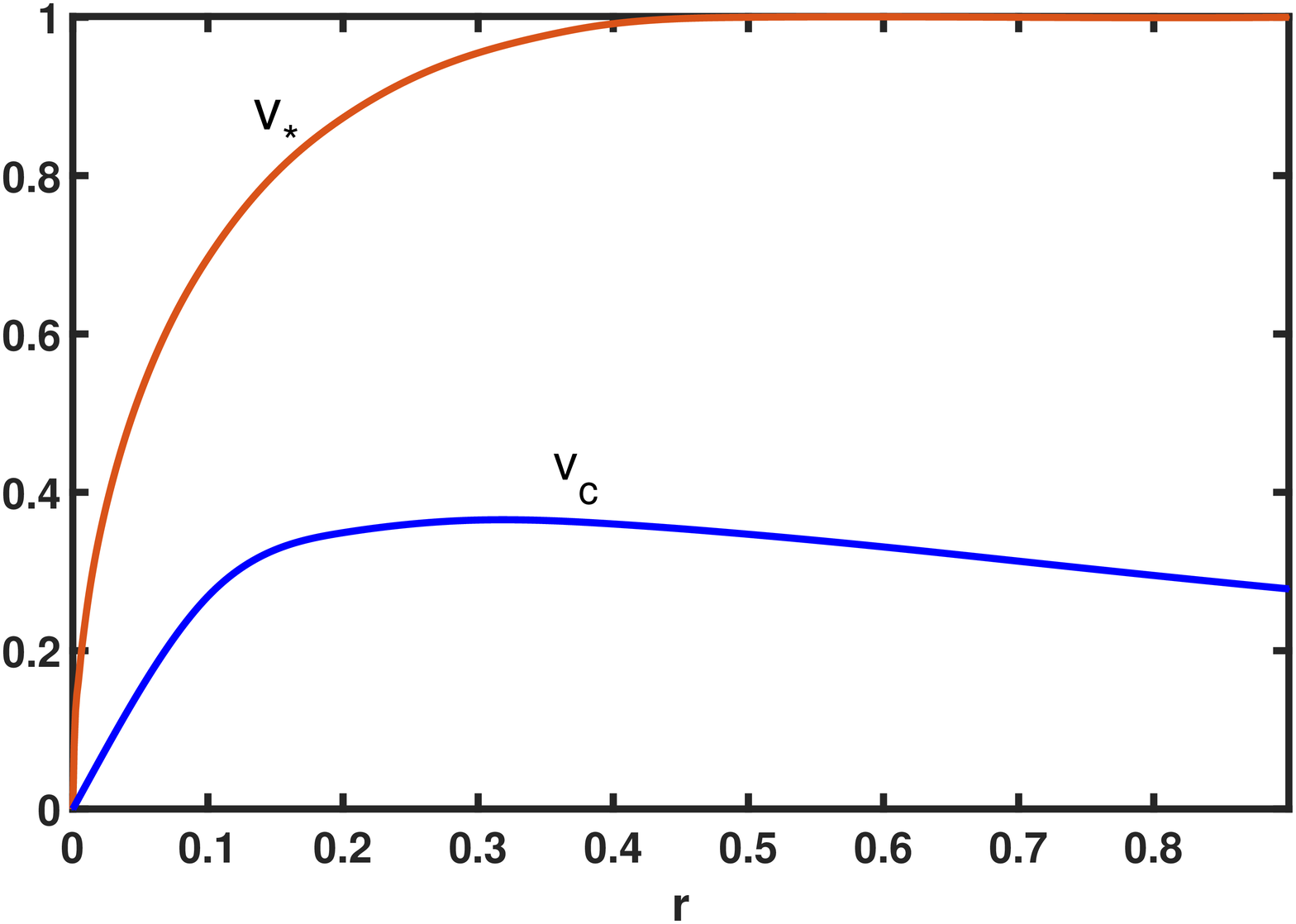}
	\caption{Small kink-antikink collision: $v_*$ and $v_c$ as a function of the parameter $r$. For $v>v_*$ there is the changing of the topological sector, with the production of large antikink-kink pair. For $v_c<v<v_*$ we have one-bounce collision. For $v<v_c$ one can have either bion or two-bounce collision. }
	\label{v*}
\end{figure}

For $v<v_*$ the kink-antikink scattering does not result in the changing of the topological sector. There are several possibilities: i) for $v_c<v<v_*$ the collision is almost elastic and produces small radiation inside the produced small kink-antikink pair. This can be seen in the Fig. \ref{small-small}a. In particular, the Fig. \ref{small-small}b shows that the scalar field at $x=0$ bounce once around the other topological sector, but returns, oscillating around the initial vacuum state. This is a one-bounce collision. ii) for $v<v_c$ one has the formation of a bion state where, in the long term, the kink-antikink pair annihilates (see the Figs. \ref{small-bion}a-b), or a two-bounce collision (see the Figs. \ref{small-2bounce}a-b).

The Fig. \ref{v*} shows the velocities $v_*$ and $v_c$ as a function of $r$. Note from the figure that $v_*$ is a monotonically growing function of $r$ and goes to zero as $r\to 0$. This agrees with the limit of the sine-Gordon model, where small kinks are absent. We also observe that the region $v>v_*$, where a large antikink-kink pair is formed, is restricted to $0<r\lesssim 0.38$. The critical velocity $v_c$ has a maximum around $r\sim0.3$.

\begin{figure}
	\includegraphics[{angle=0,width=5cm,height=4cm}]{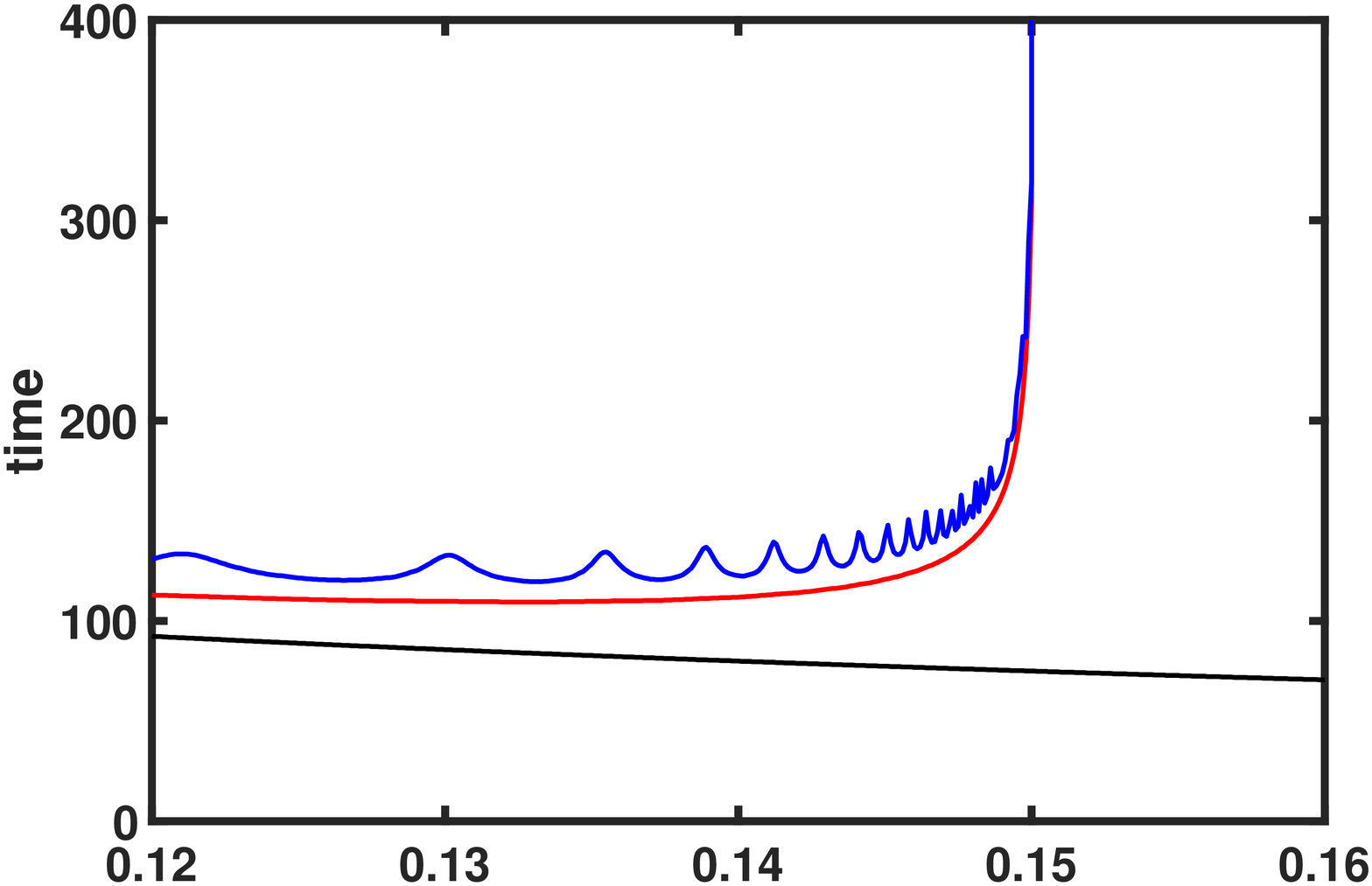}
	\includegraphics[{angle=0,width=5cm,height=4cm}]{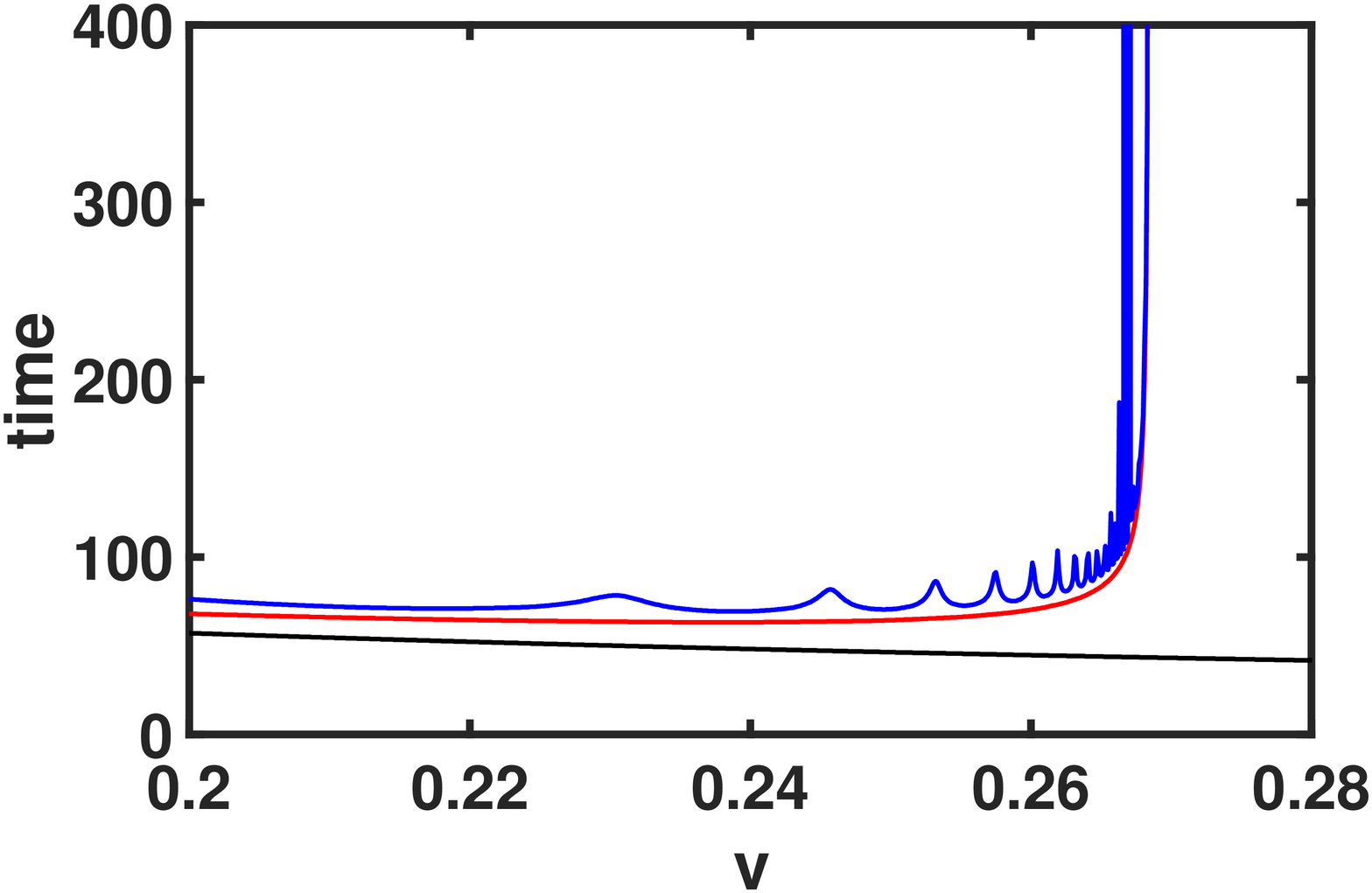}
	\includegraphics[{angle=0,width=5cm,height=4cm}]{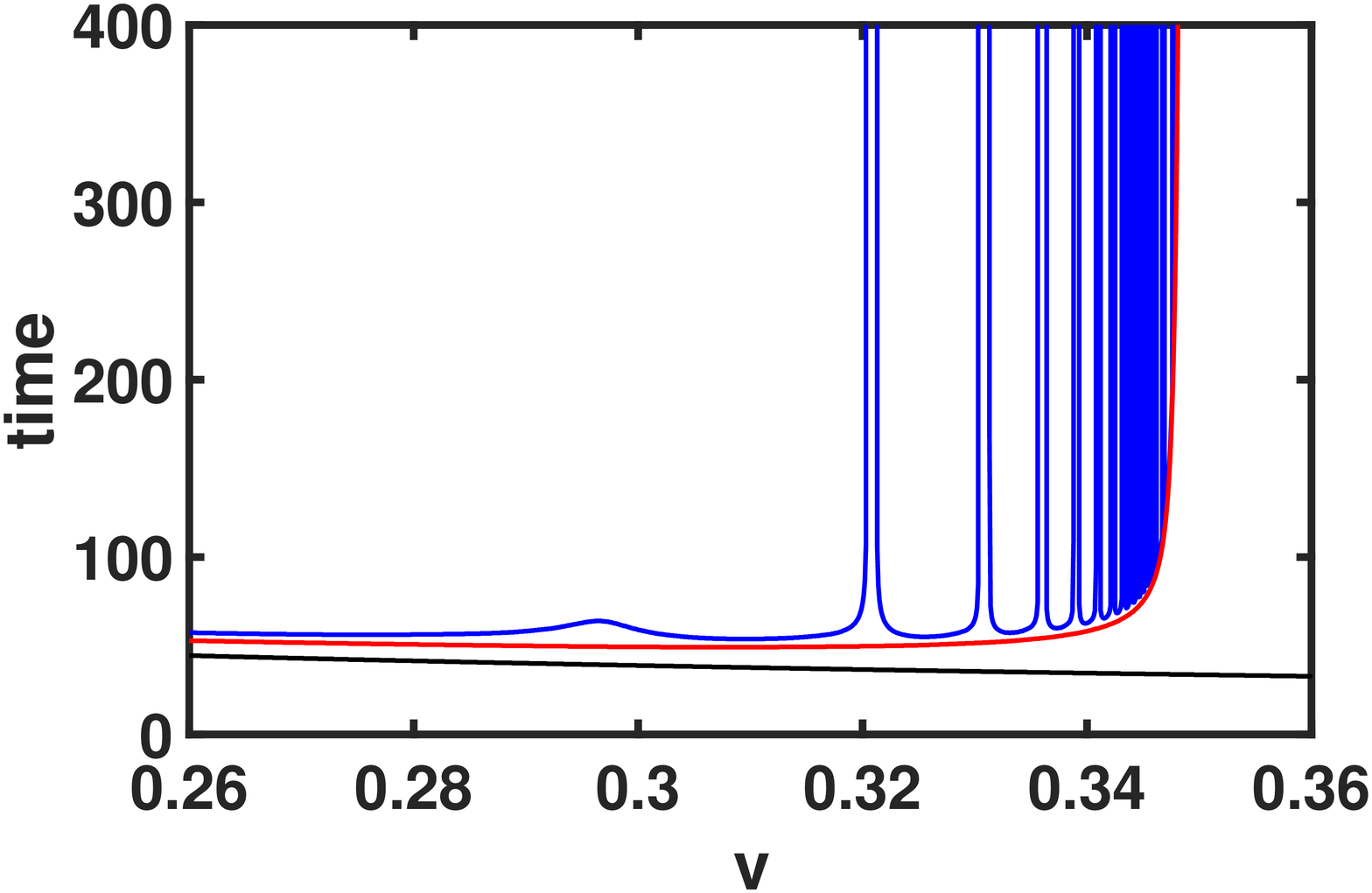}
	\caption{Small kink-antikink collision: time to first (black), second (red) and third (blue) bounces for kink-antikink colisions as a function of initial velocity for (a) $r=0.05$, (b) $r=0.1$ and (c) $r=0.2$.}
	\label{time}
\end{figure}

The structure of scattering for small kink solutions for $v<v_c$ is depicted in the Figs. \ref{time}a-c, where we show some plots of the time to first, second and third bounces for small kink-antikink collisions as a function of initial velocity for fixed values of $r$. 
In the Fig. \ref{time}a, for the case $r=0.05$, we can see the formation of bion states and false two-bounce windows (also known as quasiresonances \cite{dsg8}) for $v<v_{c}=0.151$.
Note that, despite the presence of an internal shape mode, and contrary to the expected by the resonant energy exchange mechanism \cite{csw}, there are no structures of two-bounce windows. For $r=0.1$ (Fig. \ref{time}b) we can only see the appearance of two thin two-bounce windows and the occurrence of one-bounce with $v>v_{c}=0.2685$. For $r=0.2$ (Fig. \ref{time}c), we observe the growth of critical velocity ($v_{c}=0.3486$) and the increase of quantity of two-bounce windows. In this figure there is the formation of only one false two-bounce windows. That is, the increasing of $r$ contributes to recover the full structure of two-bounce windows expected by the resonant energy exchange mechanism \cite{csw}. On the contrary, a  decreasing of $r$ shows that the two-bounce windows tend to be suppressed, with the formation of false two-bounce windows.


\section { Conclusion }


We have analyzed the double sine-Gordon model with two distinct solutions that depend on a parameter $r$. The minima of the potential are separated by large and small barriers leading respectively to large and small kink. 

It was shown that the potential of perturbations for large kink solution transits from a smooth valley to a volcano-like. This behavior leads to the appearance of resonant peaks and the absence of vibrational states. The dynamics of large kink-antikink scattering shows for small values of parameter $r$  the formation of small antikink-kink pairs. For each time, the  produced antikink-kink pairs more distant from the collision point have the smaller thickness, do not emit detected radiation and oscillates around the vacuum, with frequencies below the continuum. 
 The growth of $r$ allows the appearance of more antikink-kink pairs and even propagating oscillations. The region of $r$ where such oscillations appear to coincide with the transition region of the formation of more antikink-kink pairs.

The stability analysis of Schr\"odinger-like potential for small kink solution shows that the increase of $r$ reduces the depth of the minimum and decrease the asymptotic maximum of the potential. For small kink, the occurrence of bound states was investigated and we obtained the spectra of excitations with zero mode for all values and one vibrational mode for $r \gtrsim 0.02$. 
The scattering of small kink-antikink can result in several possibilities: i) the changing of the topological sector, producing large antikink-kink; ii) collisions without the changing of the topological sector as one-bounce or two-bounce; iii) the formation of a bion state that in the long-run annihilates the small kink-antikink pair.   
  For $r=0.05$ we observe the formation of false two-bounce windows and the suppression of structure of true two-bounce windows, despite the presence of an internal shape mode. This is an interesting result and in disagreement with the resonant energy exchange mechanism. The increasing of $r$ to large values leads to a gradual growth of the number of two-bounce windows.


\section{Acknowledgements}
We thank Giuseppe Mussardo and Harold Blas, for interesting informations concerning the double sine-Gordon model and Vakhid Gani,
for interesting suggestions and questions that improved the manuscript. We also thank FAPEMA - Funda\c c\~ao de Amparo \`a Pesquisa e ao Desenvolvimento do Maranh\~ao through grants PRONEX 01452/14, PRONEM 01852/14, Universal 01061/17, 01191/16, 01332/17, 01441/18 and BD - 00128/17. A.R.G thanks CNPq (brazilian agency) through grants 437923/2018-5 and 311501/2018-4 for financial support. This study was financed in part by the Coordena\c c\~ao de Aperfei\c coamento de Pessoal de N\'ivel Superior - Brasil (CAPES) - Finance Code 001.



\end{document}